\documentclass[11pt]{article}
\usepackage{mathrsfs}           % \mathscr
\usepackage{vmargin,fancyhdr}   % margins
\usepackage{algorithm}
\usepackage{algorithmic}

\usepackage{enumerate}
\usepackage{array}
\usepackage{amsmath,amssymb}    % AMS
\usepackage{verbatim}           % comment environment
\usepackage{xspace}             % optional space
\usepackage{graphicx,float}     % figures
\usepackage{ifthen,calc}        % macros
\usepackage{textcomp}           % \textcent
\usepackage{fancybox}           % \Ovalbox
\usepackage{hhline}             % \hhline
\usepackage{float}              % \float

\usepackage[rflt]{floatflt}     % default is vflt
\usepackage[small,compact]{titlesec}
\usepackage{setspace}
\usepackage{subfigure}

\newcolumntype{C}[1]{>{\centering\let\newline\\\arraybackslash\hspace{0pt}}m{#1}}
\usepackage{color}
\definecolor{ForestGreen}{rgb}{0.1333,0.5451,0.1333}
\usepackage[letterpaper,dvips,
            colorlinks,linkcolor=ForestGreen,citecolor=ForestGreen,
            backref,
            bookmarks,bookmarksopen,bookmarksnumbered]
           {hyperref}
\usepackage[dvips]{thumbpdf}

\setpapersize{USletter}
\setmarginsrb{.75in}{.5in}        % left, top
             {.75in}{.5in}        % right, bottom
             {.25in}{.25in}     % headheight, headsep
             {.25in}{.5in}      % footheight, footskip
\newcommand{\showccc}[0]{0}
\newcommand{\ccc}[2][nothing]{% comment
  \ifthenelse{\showccc=0}{}{
    \ensuremath{^{\Lsh\Rsh}}\marginpar{\raggedright\tiny\textsf{%
        \ifthenelse{\equal{#1}{nothing}}{}{\textbf{#1}\\}#2}}}}

\newcounter{hours}\newcounter{minutes}
\newcommand{\hhmm}{%
  \setcounter{hours}{\time/60}%
  \setcounter{minutes}{\time-\value{hours}*60}%
  \ifthenelse{\value{hours}<10}{0}{}\thehours:%
  \ifthenelse{\value{minutes}<10}{0}{}\theminutes}
\ifthenelse{\showccc=0}{\rhead{}}{\rhead{\today \ [\hhmm]}}
\newtheorem{theorem}{Theorem}[section]

\newtheorem{corollary}[theorem]{Corollary}
\newtheorem{definition}[theorem]{Definition}

\newtheorem{lemma}[theorem]{Lemma}
\newtheorem{fact}[theorem]{Fact}

\newcommand{\Proof}[0]{\smallskip\noindent\textit{\textbf{Proof}}\quad}
\newcommand{\Proofof}[1]{\smallskip\noindent\textit{\textbf{Proof of #1:}}\quad}

\newcommand{\QED}[0]{\hfill\ensuremath{\blacksquare}\medspace}

\floatstyle{ruled}
\newfloat{algo}{tbp}{lop}%[section]
\floatname{algo}{Algorithm}

\usepackage{float}
\floatstyle{plain}\newfloat{myfig}{t}{figs}[section]
\floatname{myfig}{\textsc{Figure}}
\floatstyle{plain}\newfloat{myalg}{H}{algs}[section]
\floatname{myalg}{}
\newcommand{\linearprimal}{\mathcal{OBJ}}
\newcommand{\quadprimal}{\mathcal{OBJ}2}

\newcommand{\opt}{\text{OPT}}
\newcommand{\optquad}{\text{OPT2}}

\newcommand{\fixed}{\textbf{s}}
\newcommand{\fixedv}{\textbf{s}}

\newcommand{\mata}{\textbf{A}}
\newcommand{\matb}{\textbf{B}}
\newcommand{\mident}{\textbf{I}}

\newcommand{\matc}{\textbf{C}}
\newcommand{\matu}{\textbf{U}}
\newcommand{\matv}{\textbf{V}}

\newcommand{\vecu}{\textbf{u}}
\newcommand{\vecv}{\textbf{v}}
\newcommand{\vecx}{\textbf{x}}
\newcommand{\vecy}{\textbf{y}}

\newcommand{\optvecx}{\bar{\textbf{x}}}

\newcommand{\flowv}{\textbf{f}}
\newcommand{\optflowv}{\bar{\textbf{f}}}

\newcommand{\weight}{\textbf{w}}
\newcommand{\weightv}{\textbf{w}}

\newcommand{\group}{\textbf{L}}
\newcommand{\edgevertex}{\textbf{B}}

\newcommand{\zerov}{\textbf{0}}

\newcommand{\cost}{\textbf{cost}}
\newcommand{\costv}{\textbf{cost}}

\newcommand{\weightmat}{\textbf{$\alpha$}}
\newcommand{\weightmatv}{\textbf{$\alpha$}}

\begin{document}

\title{Runtime Guarantees for Regression Problems
\thanks{Partially supported by the National Science Foundation under grant number CCF-1018463.}}

\author{
  Hui Han Chin\\
  CMU\\
%  Carnegie Mellon University\\
  \texttt{hchin@cmu.edu}\\
\and
  Aleksander M\c{a}dry\\
%  MSR New England
  EPFL\\
  \texttt{madry@mit.edu}\\
\and
  Gary L.\ Miller\\
  CMU\\
%%  Carnegie Mellon University\\
  \texttt{glmiller@cs.cmu.edu}\\
  \and
  Richard Peng
  \thanks{Supported by a Microsoft Fellowship}
  \thanks{Partially supported by Natural Sciences and Engineering Research Council of Canada (NSERC) under grant number D-390928-2010.}\\
  CMU\\
%%  Carnegie Mellon University\\
  \texttt{yangp@cs.cmu.edu}\\
%  \texttt{\{glmiller,yangp\}@cs.cmu.edu}
}

%\date{}                         % proceedings should not have date

\begin{titlepage}
	\maketitle
	\begin{abstract}
We study theoretical runtime guarantees for 
a class of optimization problems that occur in
a wide variety of inference problems.
These problems are motivated by the lasso framework
and have applications in machine learning and computer vision.

Our work shows a close connection between these problems and
core questions in algorithmic graph theory.
While this connection demonstrates the difficulties of obtaining
runtime guarantees, it also suggests an approach of
using techniques originally developed for graph algorithms.

We then show that most of these problems can be formulated as a grouped
least squares problem, and give efficient algorithms for this formulation.
Our algorithms rely on routines for solving quadratic minimization problems,
which in turn are equivalent to solving linear systems.
Finally we present some experimental results on applying our approximation
algorithm to image processing problems.
\end{abstract}

	\thispagestyle{empty}
\end{titlepage}

\section{Introduction}
\label{sec:intro}

The problem of recovering a discrete, clear signal from noisy data is an
important problem in signal processing.
One general approach to this problem is to formulate an objective based
on required properties of the answer, and then return its minimizer
via optimization algorithms.
The power of this method was first demonstrated in image denoising,
where the total variation minimization approach
by Rudin, Osher and Fatemi \cite{rof92a} had much success.
More recent works on sparse recovery led to the theory of
compressed sensing \cite{Candez06Survey}, which includes approaches
such as the least absolute shrinkage and selection operator (LASSO)
objective due to Tibshirani \cite{Tibshirani96}.
These objective functions have proven to be immensely powerful tools,
applicable to problems in signal processing, statistics, and computer vision.
In the most general form, given vector $\vecy$ and a matrix $\mata$,
one seeks to minimize:

\begin{align}
\min_{\vecx} & ||\vecy - \mata \vecx||_2^2 \\
\text{subject to:}~~~~~& |\vecx|_1 \leq c \nonumber
\label{eq:lasso}
\end{align}

It can be shown to be equivalent to the following
by introducing a Lagrangian multiplier, $\lambda$:

\begin{align}
\min_{\vecx} & ||\vecy - \mata \vecx||_2^2 + \lambda |\vecx|_1
\end{align}

Many of the algorithms used to minimize the LASSO objective in practice
are first order methods \cite{Nesterov07,BeckerCG11}, which updates
a sequence of solutions using well-defined vectors related to the gradient.
These methods are guaranteed to converge well when the
matrix $\mata$ is ``well-structured''.
The formal definition of this well-structuredness is closely related
to the conditions required by the guarantees given in the
compressed sensing literature \cite{Candez06Survey}
for the recovery of a sparse signal.
As a result, these methods perform very well on problems where
theoretical guarantees for solution quality are known.
This good performance, combined with the simplicity of implementation,
makes these algorithms the method of choice for most problems.

However, LASSO type approaches have also been successfully applied
to larger classes of problems.
This has in turn led to the use of these algorithms on a much wider
variety of problem instances.
An important case is image denoising, where works on
LASSO-type objectives predates the compressed sensing literature \cite{rof92a}.
The matrices involved here are based on the connectivity
of the underlying pixel structure, which is often a
$\sqrt{n} \times \sqrt{n}$ square mesh.
Even in a unweighted setting, these matrices tend to be ill-conditioned.
In addition, the emergence of non-local formulations that can connect arbitrary pairs of
vertices in the graph also highlights the need to handle problems that
are traditionally considered ill-conditioned.
We show in Appendix \ref{sec:graphproofs} that the broadest definition of
LASSO problems include well-studied problems from algorithmic graph theory:

\begin{fact}
\label{fac:reductionslasso}
Both the $s$-$t$ shortest path and $s$-$t$ minimum cut problems in
undirected graphs can be solved by minimizing a LASSO objective.
\end{fact}

Although linear time algorithms for unweighted shortest path
are known, finding efficient parallel algorithms for this has
been a long-standing open problem.
The current state of the art parallel algorithm for finding $1 + \epsilon$
approximate solutions, due to Cohen \cite{Cohen00}, is quite involved.
Furthermore, as the reductions done in Lemma \ref{lem:shortestpath} are readily
parallelizable, an efficient algorithm for LASSO minimization would also lead
to an efficient parallel shortest path algorithm.
This suggests that algorithms for minimizing LASSO objectives,
where each iteration involve simple, parallelizable operations, are also difficult.
Finding a minimum $s$-$t$ cut with nearly-linear running
time is also a long standing open question in algorithm design.
In fact, there are known hard instances where many algorithms do exhibit
their worst case behavior \cite{JohnsonM93}.
The difficulty of these problems and the non-linear
nature of the objective are two of the main challenges in obtaining
fast run time guarantees for grouped least squares minimization.

Previous run time guarantees for minimizing LASSO objectives rely
on general convex optimization routines \cite{BoydV04}, which take
at least $\Omega(n^{2})$ time.
As the resolution of images are typically at least $256 \times 256$,
this running time is prohibitive.
As a result, when processing image streams or videos in real time,
gradient descent or filtering based approaches are typically used
due to time constraints, often at the cost of solution quality.
The continuing increase in  problem instance size, due to higher
resolution of streaming videos, or 3D medical images with billions of voxels,
makes the study of faster algorithms an increasingly important question.

While the connection between LASSO and graph problems gives
us reasons to believe that the difficulty of
graph problems also exists in minimizing LASSO objectives, it also suggests
that techniques from algorithmic graph theory can be brought to bear.
To this end, we draw upon recent developments in algorithms for maximum
flow \cite{ChristianoKMST10} and minimum cost flow \cite{DaitchS08}.
We show that relatively direct modifications of these algorithms allows
us to solve a generalization of most LASSO objectives, which we term the
{\bf grouped least squares problem}.
Our algorithm is similar to convex optimization algorithms in that
each iteration of it solves a quadratic minimization problem,
which is equivalent to solving a linear system.
The speedup over previous algorithms come from the existence of much
faster solvers for graph related linear systems \cite{SpielmanTengSolver},
although our approaches are also applicable to situations involving
other underlying quadratic minimization problems.

The organization of this paper is as follows: In Section \ref{sec:formulations}
we provide a unified optimization problem that encompasses LASSO, fused LASSO,
and grouped LASSO.
We then discuss known applications of the grouped least squares minimization
in Section \ref{sec:applications} and existing approaches in Section \ref{sec:previous}.
In Section \ref{sec:algo} we give two algorithms that use solving quadratic
minimization problems as underlying routines: An approximate algorithm based
on the maximum flow algorithm of Christiano et al. \cite{ChristianoKMST10},
and an almost-exact algorithm that rely on interior point algorithms.

\section{Background and Formulations}
\label{sec:formulations}

The formulation of our main problem is motivated by the total
variation objective from image denoising.
This objective has its origin in the seminal work by Mumford and
Shah \cite{MumfordS89}.
There are two conflicting goals in recovering a smooth image from
a noisy one, namely that it must be close to the original image, while
having very little noise.
The Mumford-Shah function models the second constraint by imposing
penalties for neighboring pixels that differ significantly.
These terms decrease with the removal of local distortions, offsetting
the higher cost of moving further away from the input.
However, the minimization of this functional is computationally difficult
and subsequent works focused on minimizing functions that are close to it.

The total variation objective is defined for a discrete, pixel representation
of the image and measures noise using a smoothness term calculated
from differences between neighboring pixels.
This objective leads naturally to a graph $G = (V, E)$ corresponding to the image
with pixels.
The original (noisy) image is given as a vertex labeling $\fixedv$, while
the goal of the optimization problem is to recover the `true' image
$\vecx$, which is another set of vertex labels.
The requirement of $\vecx$ being close to $\fixedv$ is quantified
by $||\vecx - \fixedv||_2^2$, specifically summing over the squares
of what we identify as noise while the smoothness term is a sum over
absolute values of difference between adjacent pixels' labels:

\begin{align}
||\vecx - \fixedv||_2^2 + \sum_{(u, v) \in E} |x_u - x_v|
\label{eq:L22L1}
\end{align}

This objective can be viewed as an instance of the fused LASSO
objective \cite{TibshiraniSRZK05}.
As the orientation of the underlying pixel grid is artificially imposed
by the camera, this method can introduce rotational bias in its output.
One way to correct this bias is to group the differences
of each pixel with its 4 neighbors, giving terms of the form:

\begin{align}
\sqrt{(x_u - x_v)^2 + (x_u - x_w)^2}
\end{align}

where $v$ and $w$ are the horizontal and vertical neighbor of $u$.

Our generalization of these objectives
is based on the key observation that $\sqrt{(x_u - x_v)^2 + (x_u - x_w)^2}$
and $|x_u - x_v|$ are both $L_2$ norms of a
vector containing differences of values of adjacent pixels.
Each such difference can be viewed as an edge in the underlying graph,
and the grouping gives a natural partition of the edges
into disjoint sets $S_1 \dots S_k$:

\begin{align}
||\vecx-\fixedv||_2^2 +
\sum_{1 \leq i \leq k}
    \sqrt{\sum_{(u,v) \in S_i} (x_u - x_v)^2}
\label{eq:L22L2}
\end{align}

It can be checked that when each $S_i$ contain a single edge,
this formulation is identical to the objective in Equation \ref{eq:L22L1}
since $\sqrt{(x_u - x_v)^2} = |x_u - x_v|$.
Then all the terms can be written as quadratic positive semi-definite
terms involving $\vecx$ and $\fixedv_i$, where we now allow for
different fixed values in the groups.
Specifically when given symmetric positive semidefinite  (PSD) matrices $\group_0 \ldots \group_k$,
the objective can be rewritten as:

\begin{align}
||\vecx-\fixedv_0||_2^2 +
  \sum_{1 \leq i \leq k} \sqrt{(\vecx - \fixedv_i)^T \group_i (\vecx - \fixedv_i)}
\label{eq:L22L2matrix}
\end{align}

If we use $||\cdot||_{\group_i}$ to denote the norm induced by
the PSD matrix $\group_i$, each of the terms can be written
as $||\vecx - \fixedv_i||_{\group_i}$.
To make the first term resemble the other terms in our objective,
we will take a square root of it -- as we prove in Appendix
\ref{sec:variants}, algorithms that give exact minimizers for
this variant still captures the original version of the problem.
This simplification allows us to define our main problem:

\begin{definition}\label{def:main-prob}
The {\bf grouped least squares problem} is:

Input:
$n \times n$ matrices $\group_1 \ldots \group_k$
and fixed potentials $\fixedv_1 \ldots \fixed_k \in \Re^{n}$.

Output:
\[ \min_{\vecx} \linearprimal(\vecx) = \sum_{1 \leq i \leq k}
         ||\vecx - \fixedv_i||_{\group_i}
\]
\end{definition}

Note that this objective allows for the usual definition of LASSO
involving terms of the form $|\vecx_u|$ by having one group for each
such variable with $\fixedv = \zerov$.
It is also related to group LASSO \cite{YuanL06}, which incorporates
similar assumptions about closer dependencies among some of the terms.
To our knowledge grouping has not been studied in conjunction with fused LASSO,
although many problems such as the ones listed in Section \ref{sec:applications}
require this generalization.

\subsection{Quadratic Minimization and Solving Linear Systems}
\label{subsec:solve}

Our algorithmic approach to the group least squares problem crucially
depends on solving a related quadratic minimization problem.
Specifically, we solve linear systems involving a weighted combination
of the $\group_i$ matrices.
Let $\weight_1 \ldots \weight_k \in \Re^+$ denote weights,  where
$\weight_i$ is the weight on the $i$th group.
Then the quadratic minimization problem that we consider is:

\[ \min_{\vecx} \quadprimal(\vecx, \weightv) = \sum_{1 \leq i \leq k}
         \frac{1}{\weight_i} ||\vecx - \fixedv_i||_{\group_i}^2
\]

We will use $\optquad(\weightv)$ to denote the minimum value
that is attainable.
This minimizer, $\vecx$, can be obtained using the following Lemma:

\begin{lemma}
\label{lem:quadminimization}
$\quadprimal(\vecx, \weightv)$ is minimized for $\vecx$ such that

\begin{align}
\left( \sum_{1 \leq i \leq k} \frac{1}{\weight_i} \group_i \right) \vecx
= & \sum_{1 \leq i \leq k} \frac{1}{\weight_i} \fixedv_i
\end{align}
\end{lemma}

Therefore the quadratic minimization problem reduces to a linear system
solve involving $\sum_i \frac{1}{\weight_i} \group_i$, or
$\sum_i \weightmat_i \group_i$ where $\weightmatv$ is an arbitrary
set of positive coefficients.
In general, this can be done in $\tilde{O}(n^{\omega})$ time where
$\omega$ is the matrix multiplication constant \cite{Strassen69, CopWin90, VassilevskaWilliams12}.
When $\group_i$ is symmetric diagonally dominant, which is the case
for image applications and most graph problems, these systems
can be approximately solved to $\epsilon$ accuracy in
$\tilde{O}(m \log(1/\epsilon))$ time, where $m$ is the total number
of non-zero entries in the matrices
\cite{SpielmanTeng04, SpielmanTengSolver, KoutisMP10, KoutisMP11},
and also in $\tilde{O}(m^{1/3+\theta}\log(1/\epsilon))$ parallel depth
\cite{BlellochGKMPT11}.
There has also been work on extending this type of approach to a wider
class of systems \cite{AvronST09}, with works on systems arising from
well-spaced finite-element meshes \cite{BHV04},
2-D trusses \cite{DaitchS07}, and certain types of
quadratically coupled flows \cite{KelnerMP12}.
For the analysis of our algorithms, we treat this step as a black box with
running time $T(n, m)$.
Furthermore, to simplify our presentation we assume that the solves return
exact answers, as errors can be brought to polynomially small values with
an extra $O(\log{n})$ overhead.
We believe analyses similar to those performed in \cite{ChristianoKMST10,
KelnerMP12} can be adapted if we use approximate solves instead of
exact ones.

\section{Applications}

\label{sec:applications}

A variety of problems ranging from computer vision to statistics can be
formulated as grouped least squares.
We describe some of them below, starting with classical problems from
image processing.

\subsection{Total Variation Minimization}
\label{subsec:tv}

As mentioned in Section \ref{sec:intro}, one of the earliest applications
of these objectives was in the context of image processing.
More commonly known as total variation minimization in this setting \cite{chan05book},
various variants of the objective have been proposed with the
anisotropic objective the same as Equation \ref{eq:L22L1} and the
isotropic objective being the one shown in Equation \ref{eq:L22L2}.

Obtaining a unified algorithm for isotropic and anisotropic TV was one of the
main motivations for our work.
Our result readily applies to both situations, giving approximate algorithms
that run in $\tilde{O}(m^{4/3}\epsilon^{-8/3})$ for both cases.
It's worth noting that this guarantee does not rely on the underlying
structure of the of the graph.
This makes the algorithm readily applicable to 3-D images or non-local
models involving the addition of edges across the image .
However, when the neighborhoods are those of a 2-D image,
a saving by a $\log{n}$ factor can be obtained by using the optimal
solver for planar systems by Koutis and Miller \cite{KoutisMi07}.

\subsection{Denoising with Multiple Colors}
\label{subsec:colors}

Most works on image denoising deals with images where each pixel is described
using a single number corresponding to its intensity.
A natural extension would be to colored images, where each pixel has a set of $c$ attributes
(in the RGB case, $c=3$).
One possible analogue of $|x_i-x_j|$ in this case would be $||x_i-x_j||_2$, and this modification
can be incorporated by replacing a cluster involving a single edge with clusters over the
 $c$ edges between the corresponding pixels.

This type of approach can be viewed as an instance image reconstruction
algorithms using Markov random fields.
Instead of labeling each vertex with a single attribute, a set of $c$ attributes are used
instead and the correlation between vertices is represented using arbitrary PSD matrices.
It's worth remarking that when such matrices have bounded condition number,
it was shown in \cite{KelnerMP12} that the
resulting least squares problem can still be solved efficiently by preconditioning
with SDD matrices, yielding a similar overall running time.

\subsection{Poisson Image Editing}
\label{subsec:poisson}
The Poisson Image Editing method of Perez, Gangnet and Blake \cite{Perez03a}
is a very popular method for image blending.
This method aims to minimize the difference between the gradient of the
image and a guidance field vector $\vecv$.
We show here that the grouped least square problem can be used for minimizing
objectives from this framework.
The objective function given in equation (6) of \cite{Perez03a}

\[
\min_{f| \Omega} \sum_{(p,q) \cap \Omega \neq \emptyset } (f_p - f_q - v_{pq})^2, 
\;\;\text{with } f_p = f^*_p, \forall p \in \partial \Omega
\]
comprises mainly of terms of the form:

\begin{align*}
(x_p - x_q - v_{pq})^2
\end{align*}

This term can be rewritten as $((x_p - x_q) - (v_{pq}-0)  )^2$.
So if we let $\fixedv_i$ be the vector where $s_{i,p} = v_{pq}$ and
$s_{i,q} = 0$, and $\group_i$ be the graph Laplacian for
the edge connecting $p$ and $q$, then the
term equals to $||\vecx - \fixedv_{i}||^2_{\group_i}$.
The other terms on the boundary will have $x_q$ as a constant,
leading to terms of the form $||x_{i,p} - s_{i,p}||^2_2$ where $s_{i,p} = x_q$.
Therefore the discrete Poisson problem of minimizing the sum of these
squares is an instance of the quadratic minimization problem as
described in Section \ref{subsec:solve}.
Perez et al. in Section 2 of their paper observed that these linear systems are
sparse, symmetric and positive definite.
We make the additional observation here that the systems involved are also
symmetric diagonally dominant.
The use of the grouped least squares framework also allows the possibility
of augmenting these objectives with additional $L_1$ or $L_2$ terms.

\subsection{Clustering}
\label{subsec:clustering}

Hocking et al. \cite{HockingVBJ11} recently studied an approach for clustering points
in $d$ dimensional space.
Given a set of points $x_1 \ldots x_n \in \Re^{d}$, one method that they proposed
is the minimization of the following objective function:

\[
\min_{y_1 \ldots y_n \in \Re^{d}}
\sum_{i = 1}^n ||x_i - y_i||_2^2
+ \lambda \sum_{ij} w_{ij} ||y_i - y_j||_2
\]

Where $w_{ij}$ are weights indicating the association between items $i$ and $j$.
This problem can be viewed in the grouped least squares framework by
viewing each $x_i$ and $y_i$ as a list of $d$ variables, giving that
the $||x_i - y_i||_2$ and $||y_i - y_j||_2$ terms can be represented using
a cluster of $d$ edges.
Hocking et al. used the Frank-Wolfe algorithm to minimize a relaxed form of this
objective and observed fast behavior in practice.
In the grouped least squares framework, this problem is an instance with
$O(n^2)$ groups and $O(dn^{2})$ edges.
Combining with the observation that the underlying quadratic optimization
problems can be solved efficiently allows us to obtain an 
$1+\epsilon$ approximate solution in $\tilde{O}(dn^{8/3}\epsilon^{-8/3})$ time.

\section{Previous Algorithmic Results}
\label{sec:previous}

Due to the importance of optimization problems motivated by LASSO
there has been much work on efficient algorithms for them.
We briefly describe some of the previous approaches for
LASSO minimization below.

\subsection{Second-order Cone Programming}
To the best of our knowledge, the only algorithms that provide robust worst-case
bounds for the entire class of grouped least squares problems are based on
applying tools from convex optimization.
In particular, it is known that interior point methods applied to these problems converge in $\tilde{O}(\sqrt{k})$ iterations with each iterations requiring solving a certain linear system \cite{BoydV04, GoldfarbY04}.
Unfortunately, computing these solutions is computationally expensive -- the best
previous bound for solving one of these systems is $O(m^{\omega})$ where
$\omega$ is the matrix multiplication exponent.
This results in fairly large $O(m^{1/2+\omega})$ total running time and contributes
to the popularity of first-order methods described above in practical scenarios.
We will revisit this approach in Subsection \ref{subsec:exact} and show an improved
algorithm for the inner iterations.
However, its running time still has a fairly large dependency on $k$.

\subsection{Graph Cuts}
For the anisotropic total variation objective shown in Equation \ref{eq:L22L1},
a minimizer can be found by solving a large number of
almost-exact maximum flow calls \cite{DarbonSigelle06a, KolmogorovZabih04}.
Although the number of iterations can be large, these works show that
the number of problem instances that a pixel can appear in is small.
Combining this reduction with the fastest known exact algorithm for the
maximum flow problem by Goldberg and Rao \cite{GoldbergR98} gives an
algorithm that runs in $\tilde{O}(m^{3/2})$ time. 

It's worth mentioning that both of these algorithms requires extracting
the minimum cut in order to construct the problems for subsequent
iterations.
As a result, it's not clear whether recent advances on fast approximations
of maximum flow and  minimum $s$-$t$ cuts \cite{ChristianoKMST10} can
be used as a black box with these algorithms.
Extending this approach to the non-linear isotropic objective also appears
to be difficult.

\subsection{Iterative Reweighted Least Squares}
An approach similar to convex optimization methods, but has
much better observed rates of convergence is the iterative reweighted least
squares (IRLS) method.
This method does a much more aggressive adjustment each iteration
and to give good performances in practice \cite{WohlbergRodriguez07}.

\subsection{First Order Methods}
The method of choice in practice are first order methods such as
\cite{Nesterov07,BeckerCG11}.
Theoretically these methods are known to converge rapidly when the
objective function satisfies certain Lipschitz conditions.
Many of the more recent works on first order methods focus on
lowering the dependency of $\epsilon$ under these conditions.
As discussed in Section \ref{sec:intro} and Appendix \ref{sec:graphproofs},
this direction can be considered orthogonal to our guarantees
as the grouped least squares problem is a significantly more general formulation.

\section{Solving Grouped Least Squares Using Quadratic Minimizations}
\label{sec:algo}

In this section, we show two algorithms for the grouped least squares problem
based on direct adaptations of state-of-the-art algorithms for maximum
flow \cite{ChristianoKMST10} and minimum cost flow \cite{DaitchS08}.
Our guarantees can be viewed as reductions to the quadratic minimization
problems described in Section \ref{subsec:solve}.
As a result, they imply efficient algorithms for problems where
fast algorithms are known for the corresponding least squares problems.
The analyses of these algorithms are intricate, but mostly follow
the presentations in \cite{ChristianoKMST10, DaitchS08, BoydV04}.
They're presented in Appendices \ref{sec:multiplicativeweights} and
\ref{sec:hessian}.

\subsection{Approximate Algorithm}
\label{subsec:approx}

Our first algorithm is based on the electrical flow based maximum flow and minimum
cut algorithm by Christiano et al. \cite{ChristianoKMST10}.
Recall that the minimum $s$-$t$ cut problem - equivalent to an
$L_1$-minimization problem - is a special case of the grouped least squares
problem where each edge belongs to its own group(i.e., $k=m$).
As a result, it's natural to extend the approach of
\cite{ChristianoKMST10} to the whole spectrum of values of $k$
by treating each group as an 'edge'.

One view of the cut algorithm from \cite{ChristianoKMST10} is that it
places a weight on each group, and minimizes a quadratic, or $L_2^2$
problem involving terms of the from
$\frac{1}{\weight_i} ||\vecx - \fixedv_i||_{\group_i}^2$.
These weights are adjusted using the multiplicative weights update
framework \cite{AroraHK05, LittlestoneW94} based on the energy
of each group.
The terms $||\vecx - \fixedv_i||_{\group_i}$ and
$\frac{1}{\weight_i} ||\vecx - \fixedv_i||_{\group_i}^2$
are equal when $\weight_{i} =||\vecx - \fixedv_i||_{\group_i}$.
Therefore, one natural view of this routine is that it gradually
adjusts the weights to become a scaled copy of
$||\optvecx - \fixedv_i||_{\group_i}$.
This leads to a simplification of the Christiano et al. algorithm, whose
update requires a flow obtained from the dual of the
quadratic minimization problem.
Pseudocode of the algorithm is shown in Algorithm \ref{alg:algoapprox}.

\begin{algorithm}[ht]
\qquad

\textsc{ApproxGroupedLeastSquares}
\vspace{0.05cm}

\underline{Input:}
PSD matrices $\group_1 \ldots \group_k$,
fixed values $\fixedv_1 \ldots \fixedv_k$ for each group.
Routine $\textsc{Solve}$ for solving linear systems,
width parameter $\rho$ and error bound $\epsilon$.

\underline{Output:} Vector $\vecx$ such that $\linearprimal(\vecx) \leq (1+10 \epsilon)\opt$.

\vspace{0.2cm}

\begin{algorithmic}[1]
\STATE{Initialize $\weight^{(0)}_i = 1$ for all $1 \leq i \leq k$}
\STATE{$N \leftarrow 10 \rho \log{n} \epsilon^{-2}$}
\FOR{$t = 1 \ldots N$}
	\STATE{$\mu^{(t-1)} \leftarrow \sum_i \weight^{(t-1)}_i$} \label{ln:defmu}
	\STATE{Use \textsc{Solve} to compute a minimizer for the quadratic minimization problem where $\alpha_i = \frac{1}{\weight^{(t - 1)}_i}$, let this solution be $\vecx^{(t)}$}
	\STATE{Let $\lambda^{(t)} = \sqrt{\mu^{(t - 1)} \quadprimal(\vecx^{(t)})}$ }
\label{ln:lambda}
	\STATE{Update the weight of each group: $\weight^{(t)}_i \leftarrow
\weight^{(t-1)}_i +\left( \frac{\epsilon}{\rho} \frac{||\vecx^{(t)} - \fixedv_i||_{\group_i}}{\lambda^{(t)}} + \frac{2 \epsilon^2}{k\rho} \right) \mu^{(t-1)}$} \label{ln:update}
\ENDFOR
\STATE{$\bar{t} \leftarrow \arg\min_{0 \leq t \leq N} \linearprimal(\vecx^{(t)})$}
\RETURN{$\vecx^{(\bar{t})}$}
\end{algorithmic}

\caption{Algorithm for the approximate decision problem of whether there exist vertex potentials with objective at most $\opt$}

\label{alg:algoapprox}
\end{algorithm}

The main difficulty of analyzing this algorithm is that the analysis of
minimum $s$-$t$ cut algorithm of \cite{ChristianoKMST10} relies
strongly on the existence of a solution where $\vecx$ is either $0$ or $1$.
Our analysis extends this potential function into the fractional
setting via. a function based on the Kulback-Liebler (KL) divergence
\cite{KullbackL51}.
To our knowledge the use of this potential with multiplicative weights
was first introduced by Freund and Schapire \cite{FreundS99}, and
is common in learning theory.
This function can be viewed as measuring the KL-divergence between
$\weight_i^{(t)}$ and $||\optvecx - \fixedv_i||_{\group_i}$ over all groups,
where $\optvecx$ an optimum solution to the grouped least squares problem.
Formally, the KL divergence between these two vectors is:

\begin{align}
D_{KL} = \sum_i || \optvecx - \fixedv_i ||_{\group_i}
        \log \left( \frac{||\optvecx - \fixedv_i||_{\group_i}}{\weight_i^{(t)}} \right) \label{eq:nudef}
\end{align}

Subtracting  the constant term given by
$\sum_i || \optvecx - \fixedv_i ||_{\group_i^{(t)}}
\log(||\optvecx - \fixedv_i||_{\group_i^{(t)}})$
and multiplying by $-1/\opt$ gives us our key
potential function, $\nu^{(t)}$:

\begin{align}
\nu^{(t)} = \frac{1}{\opt} \sum_i || \optvecx - \fixedv_i ||_{\group_i}
        \log(\weight_i^{(t)}) \label{eq:nudef}
\end{align}

It's worth noting that in the case of the cut algorithm, this
function is identical to the potential function used in
\cite{ChristianoKMST10}.
We show the convergence of our algorithm by proving that
if the solution produced in some iteration is
far from optimal, $\nu^{(t)}$ increases substantially.
Upper bounding it with a term related to the sum of
weights, $\mu^{(t)}$ allows us to prove convergence.
The full proof is given in Appendix \ref{sec:multiplicativeweights}.

To simplify the analysis, we assume that the guess that we're trying to solve the
decision problem on, $\opt$, all entries of $\fixedv$,
and spectrum of $\group_i$ are polynomially bounded in $n$.
That is, there exist some constant $d$ such that $-n^{d} \leq \fixed_{i, u} \leq n^d$
and $n^{-d} \mident \preceq \sum_i \group_i \preceq n^{d} \mident$ where
$\mata \preceq \matb$ means $\matb - \mata$ is PSD.
Some of these assumptions can be relaxed via. analyses
similar to Section 2 of \cite{ChristianoKMST10}.

\begin{theorem}
\label{thm:algoapprox}
On input of an instance of $\linearprimal$ with edges partitioned into $k$ sets.
If all parameters polynomially bounded between $n^{-d}$ and $n^d$,
running \textsc{ApproxGroupedLeastSquares} with $\rho = 2 k^{1/3} \epsilon^{-2/3}$
returns a solution $\vecx$ with such that
$\linearprimal(\vecx) \leq \max \{(1+10\epsilon)\opt,  n^{-d} \}$
where $\opt$ is the value of the optimum solution.
\end{theorem}

The additive $n^{-d}$ case is included to deal with
the case where $\opt = 0$, or is close to it. 
We believe it should be also possible to handle this case
via. condition numbers restrictions on $\sum_{i} \group_i$.

\subsection{Almost-Exact Algorithm}
\label{subsec:exact}

We now show improved algorithms for solving the second order cone
programming formulation given in \cite{BoydV04, GoldfarbY04}.
It was shown in \cite{DaitchS08} that in the linear case, as with
graph problems such as maximum flow, minimum cost flow and shortest path,
interior point algorithms reduce the problem to solving
$\tilde{O}(m^{1/2})$ symmetrically diagonally dominant linear systems.
The grouped least squares formulation creates artifacts that perturb
the linear systems generated by the interior point algorithms,
making the resulting system both more difficult to interpret
and to solve.
However, the iteration count of this approach also only depends on
$k$, \cite{GoldfarbY04,BoydV04},
and has a better dependency on $\epsilon$ of $O(\log(1/\epsilon))$.

There are various ways to solve the grouped least squares
problem using interior point algorithms.
We follow the log-barrier method, as presented in Boyd
and Vandenberghe \cite{BoydV04} here for simplicity.
This formulation defines one extra variable $y_i$ for each group
and enforces $y_i \geq ||\vecx - \fixedv_i||_{\group_i}$ using the
barrier function $\phi_i(\vecx, y_i) = \log(y_i^2 - ||\vecx - \fixedv_i||_{\group_i}^2)$.
Minimizing $t \cdot (\sum_i y_i)$ for gradually increasing values of
$t$ gives the following sequence of functions to minimize:

\begin{align}
f(t, \vecx, \vecy) = t \sum_i y_i - \sum_i \log(y_i^2 - ||\vecx - \fixedv_i||_{\group_i}^2)
\label{eq:logbarrier}
\end{align}

Various interior point algorithms have been proposed,
one commonality that they have is finding an update direction
by solving a linear system.
The iteration guarantees for recovering almost exact
solution can be characterized as follows:

\begin{lemma}
\label{lem:interiorpoint}
(Section 11.5.3 from \cite{BoydV04})
A solution that's within additive $\epsilon$ of the optimum solution
can be produced in $\tilde{O}(k^{1/2}\log(1/\epsilon))$ steps,
each of which requires solving a linear system involving
$\nabla^2 f(t, \vecx, \vecy)$ for some value of $t$, $\vecx$ and $\vecy$.
\end{lemma}

These systems are examined in detail in Appendix \ref{sec:hessian}.
Since the $t \sum_i y_i$ term is linear, it can be omitted from the
Hessian, leaving $\sum_i \nabla^2 \phi(\vecx, y_i)$.
We then check that the barrier term $y_i$ creates a low rank
perturbation to the $\group_i$ term, which is the Hessian for
$||\vecx - \fixedv_i||_{\group_i}^2$.
By taking Schur complements and applying the
Sherman-Morrison-Woodbury identity on inverses for low
rank perturbations, we arrive at the following observation
in Appendix \ref{sec:hessian}:

\begin{theorem}
\label{thm:hessiansolve}
Suppose there is an algorithm for solving linear systems of the form
$\sum_i \alpha_i \group_i$ in $T(n, m)$ time where $m$.
For any choice of $\vecx, \vecy$, a linear system involving the Hessian
of $\phi(\vecx, \vecy)$, $\nabla^2 \phi(\vecx, \vecy)$ can be solved 
in $O(k^{\omega} + k T(n, m) + k^2n)$ time.
\end{theorem}

Combining this with the iteration count of $\tilde{O}(k^{1/2}\log(1/\epsilon))$
gives a total running time of $\tilde{O}((k^{\omega + 1/2} + k^{3/2}T(n, m)
+ k^{5/2}n) \log(1/\epsilon))$.

\section{Experimental Results}
\label{sec:experiments}

We performed a series of experiments using the approximate algorithm
described in Section \ref{subsec:approx}.
The SDD linear systems that arise in the quadratic minimization problems were
solved using the combinatorial multigrid (CMG) solver \cite{KoutisMiller09, KoutisMT09}.
One side observation from our experiments is that for the
sparse SDD linear systems that arise from image processing,
the CMG solver yields good results both in accuracy and running time.

\subsection{Total Variational Denoising}
\label{subsec:TV}
\begin{figure}[H]
\begin{center}
\begin{tabular}{cccc}
Noisy Version &
Goldstein-Osher \cite{GoldsteinOsher09} &
Micchelli et al. \cite{MicchelliSX11} &
Grouped Least Squares
\\
\includegraphics[width = 41mm, height = 32mm]{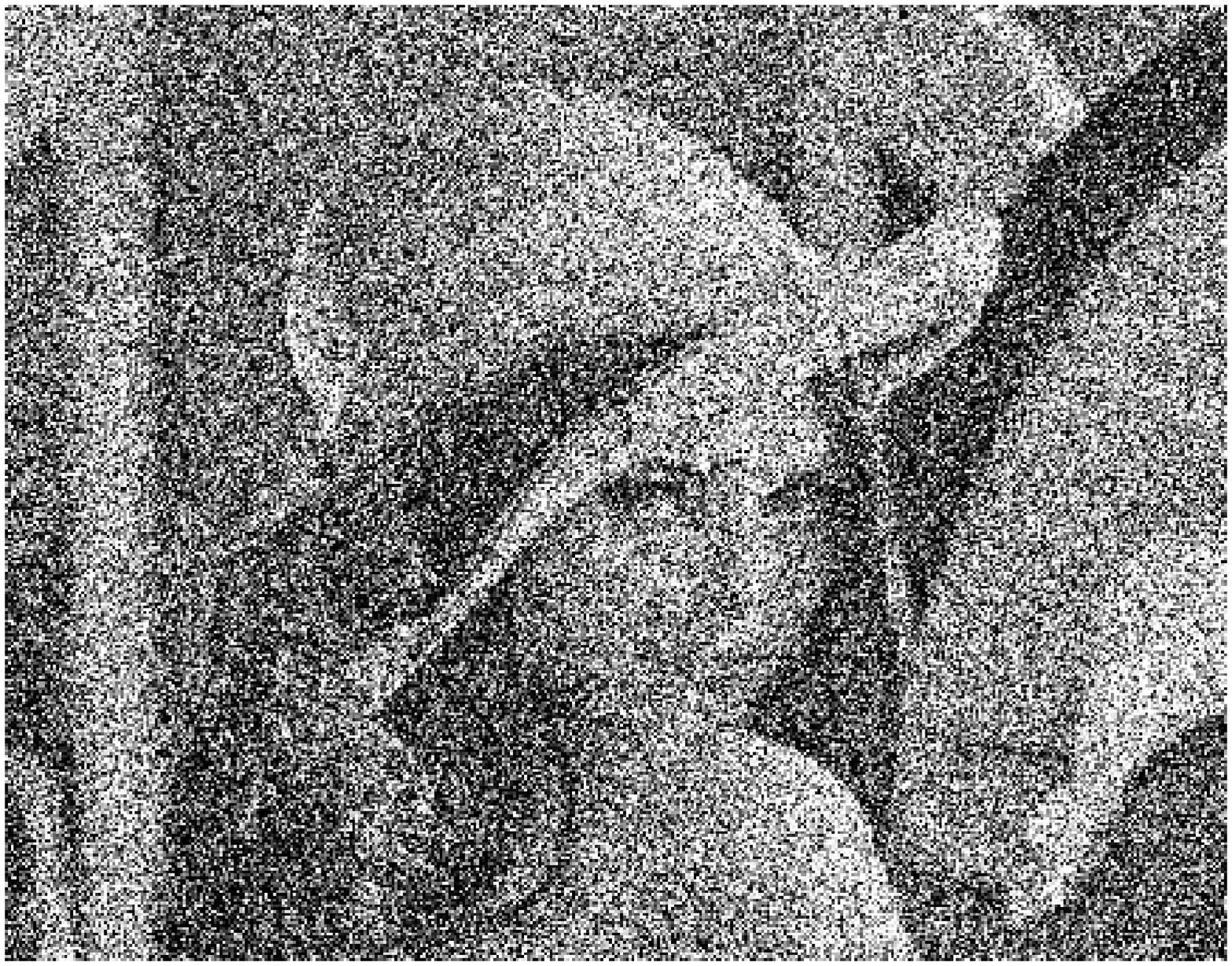}& 
\includegraphics[width = 41mm, height = 32mm]{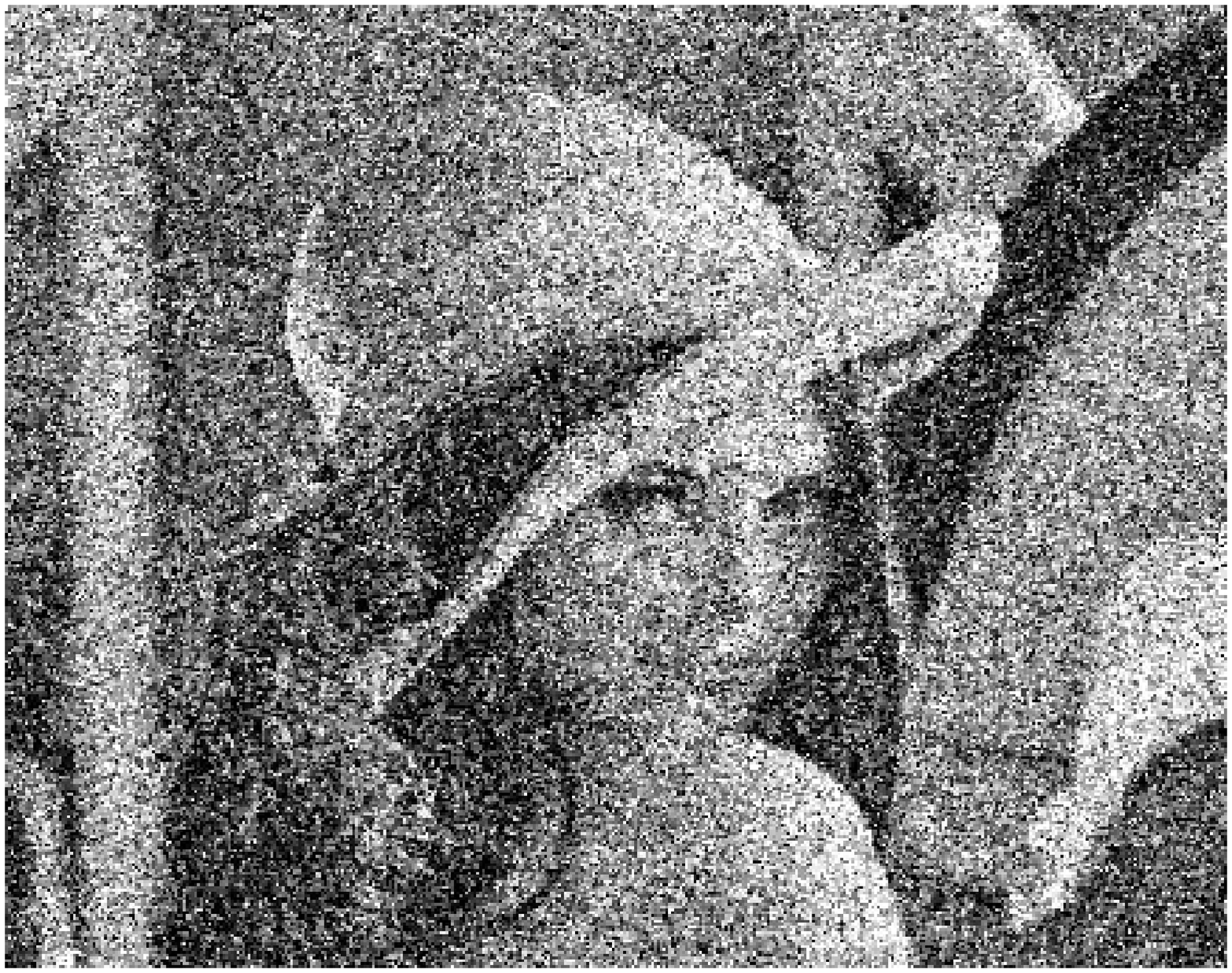} &
\includegraphics[width = 41mm, height = 32mm]{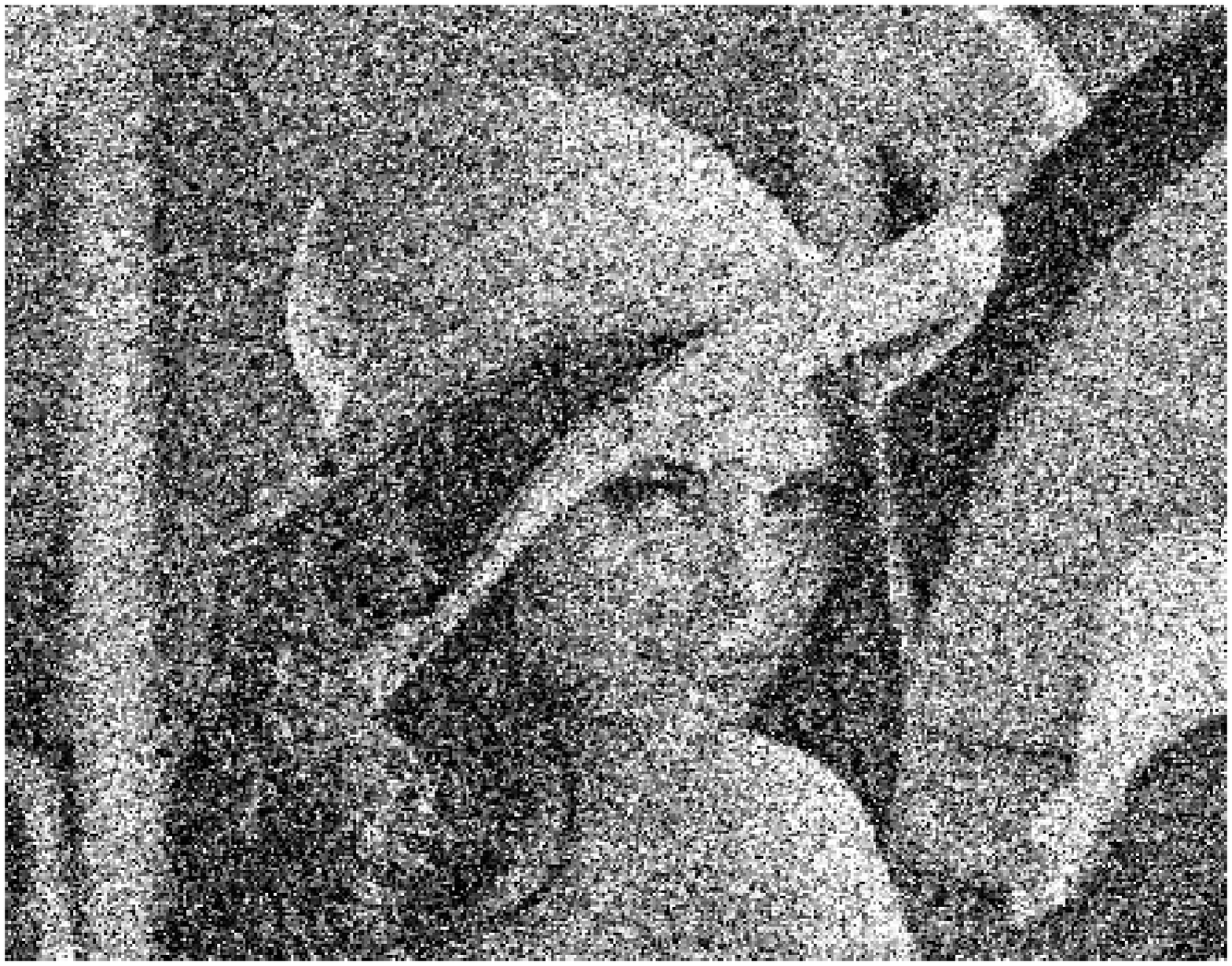} &
\includegraphics[width = 41mm, height = 32mm]{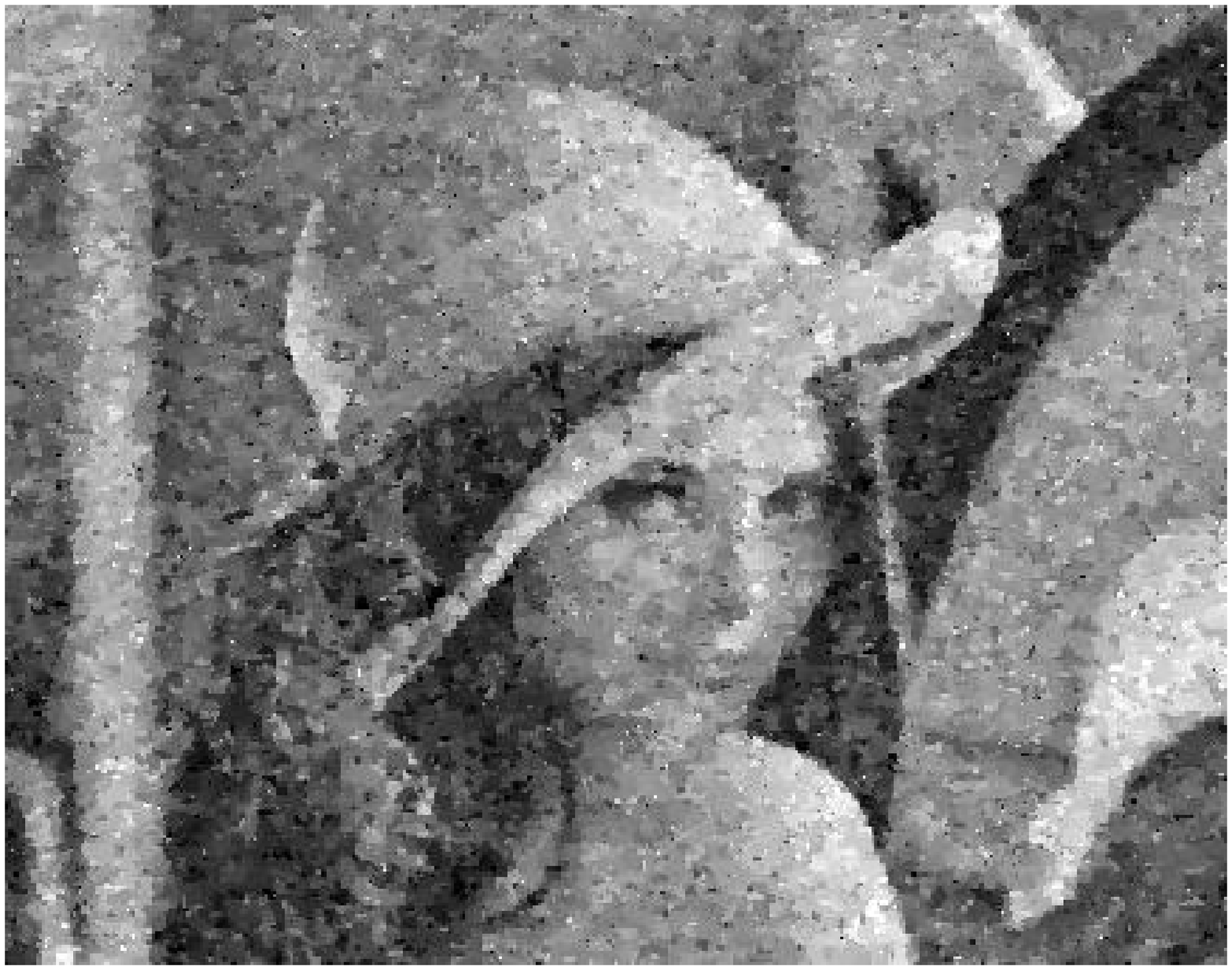} \\
22.5E6 &
13.0E6 &
14.6E6&
3.89E6
\end{tabular}
\end{center}

\caption{Outputs of Various Algorithms for Denoising Image with AWGN Noise
with Total Error Below}
\label{fig:leena}
\end{figure}

Total Variational Denoising is the concept of applying Total
Variational Minimization as denoising process.
This was pioneered by Rudin, Osher and Fatemi \cite{rof92a}
and is commonly known as the ROF image model \cite{chan05book}.
Our algorithm from Section \ref{subsec:approx} yields a simple
way to solve the ROF model and most of its variants.
In Figure \ref{fig:leena}, we present a simple denoising experiment
using the standard image processing data set, `Leena'.
The main goal of the experiment is to show that our algorithm is
competitive in terms of accuracy, while having running times
comparable to first-order methods.
On a $512 \times 512$ grayscale image, we introduce Additive White
Gaussian Noise (AWGN) at a measured  Signal to Noise Ratio (SNR) of 2.
AWGN is the most common noise model in photon capturing sensors from 
consumer cameras to space satellites systems.
Error is measured as the intensity difference from the original,
uncorrupted image summed over all pixels

Our experiments were conducted on a single core 64-bit Intel(R) Xeon(R)
E5440 CPU @ 2.83GHz.
The non-solver portion of the algorithm was implemented in Matlab(R).
On images of size $256 \times 256$, $512 \times 512$ and $1024 \times 1024$,
the average running times are $2.31$, $9.70$ and $47.61$ seconds respectively.

It's worth noting is that on average 45\% of the total running time
is from solving the SDD linear systems using the CMG solver.
The rest is due to reweighting edges in Matlab, and should be handled
as part of the CMG solver routine in more optimized versions. 
More importantly, in all of our experiments the weights are
observed to converge in under 15 iterations,
even for larger images of size up to $3000 \times 3000$.
This is much better than the guarantees given for either
of our algorithms in Section \ref{sec:algo}.

\subsection{Image Processing}
\label{subsec:imageprocessing}

As exemplified by the denoising with colors application in
Section \ref{subsec:colors}, the grouped least squares framework
has great flexibility in expressing image processing tasks.
We applied our denoising algorithm to Optical Coherence Tomography
(OCT) images of the retina as a preprocessing step for segmentation.
Here the key is to preserve the sharpness between the nerve fiber
layers and this is achieve by using a $L_1$ regularization term.

Variations of this formulation allows one to model a large variety
of established image preprocessing applications.
For example, Gaussian blurred images can be obtained
using a $L_2$ penalty term.
This simulates camera zoom and is similar to the preprocessing
step in the popular Scale Invariant Feature Transform (SIFT) algorithm.
By mixing and matching penalty terms on the groups, 
we can preserve global features while favoring the removal of small
artifacts that are often the result of sensor noise.
Our approaches also extend naturally to multichannel images,
(RGB or multi-spectral), with little modification to the
underlying algorithm.

\begin{figure}[H]
\begin{center}
\begin{tabular}{*{3}{C{5.5cm}}}
\includegraphics[width=50mm, height=50mm]{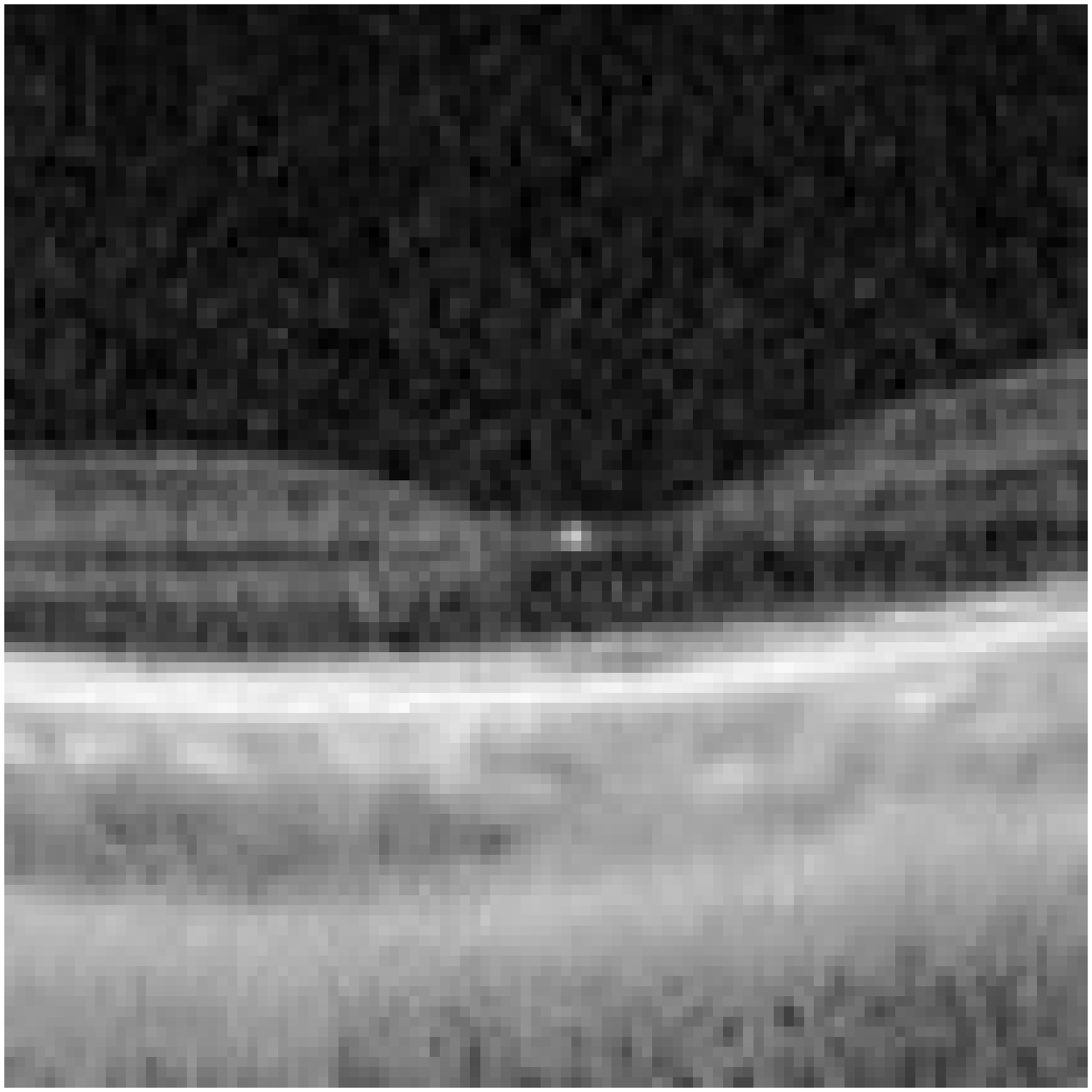}
& \includegraphics[width=50mm, height=50mm]{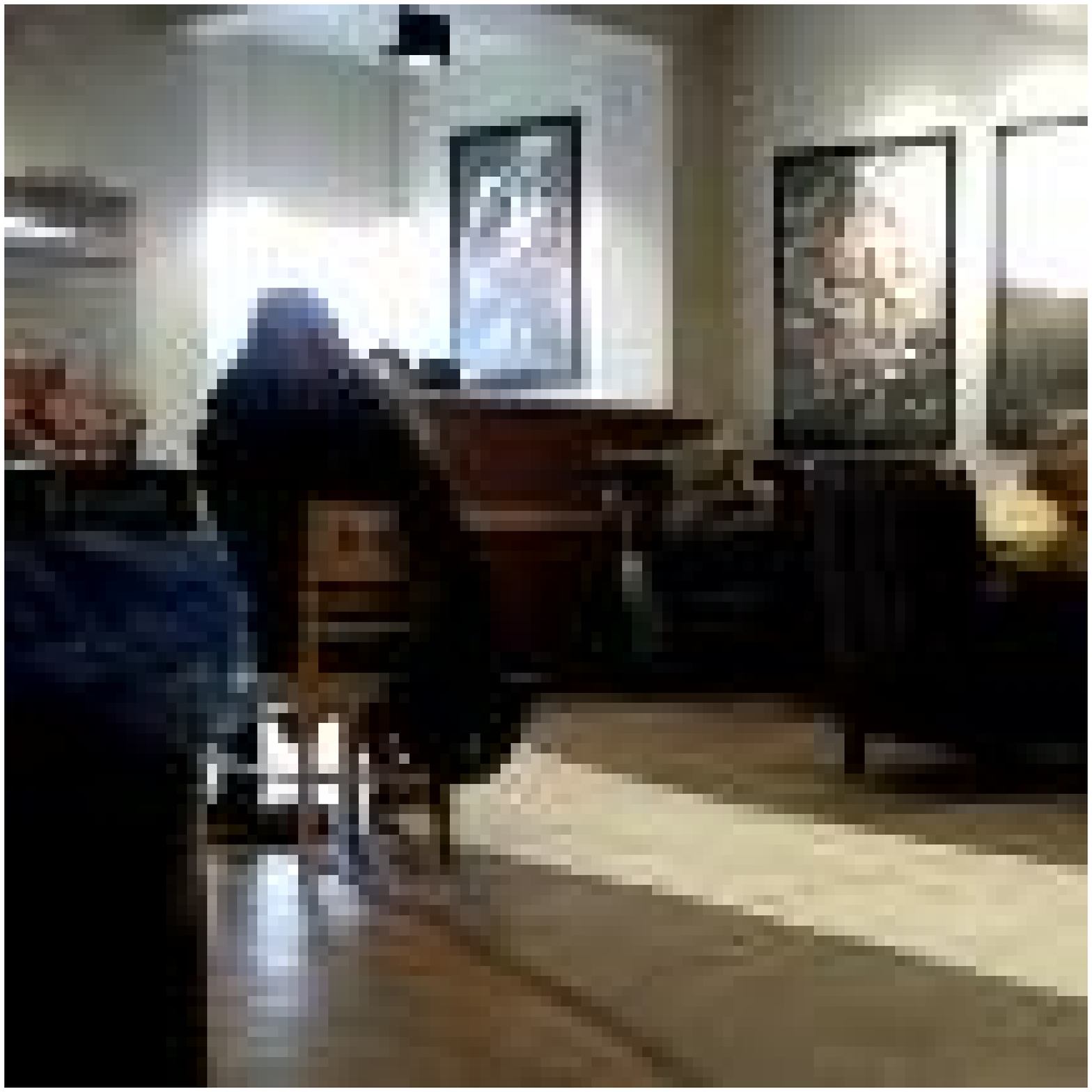}
& \includegraphics[width=50mm, height=50mm]{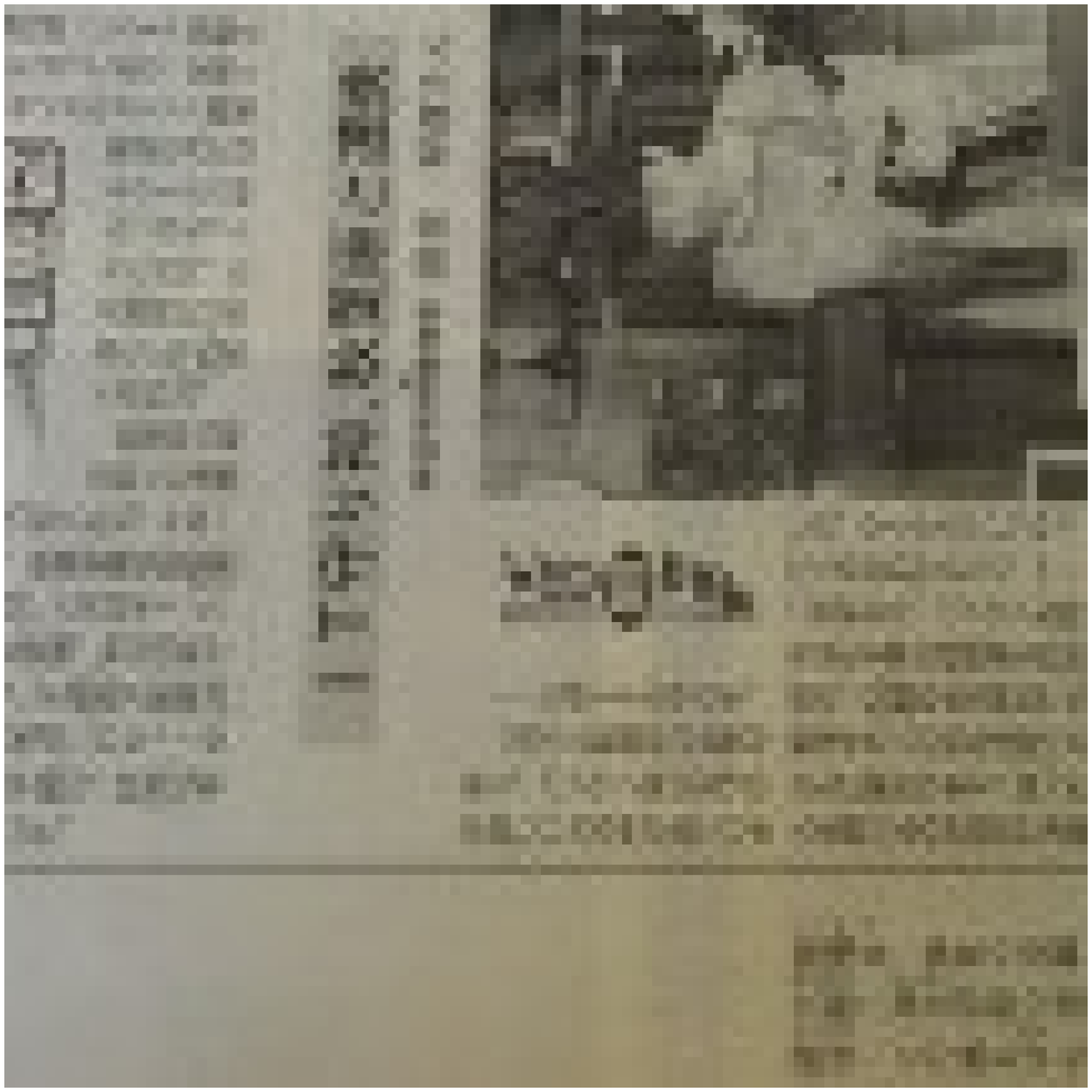} \\
Noisy OCT image of retina
& Color image
&  Yellowed newspaper scan \\
\includegraphics[width=50mm, height=50mm]{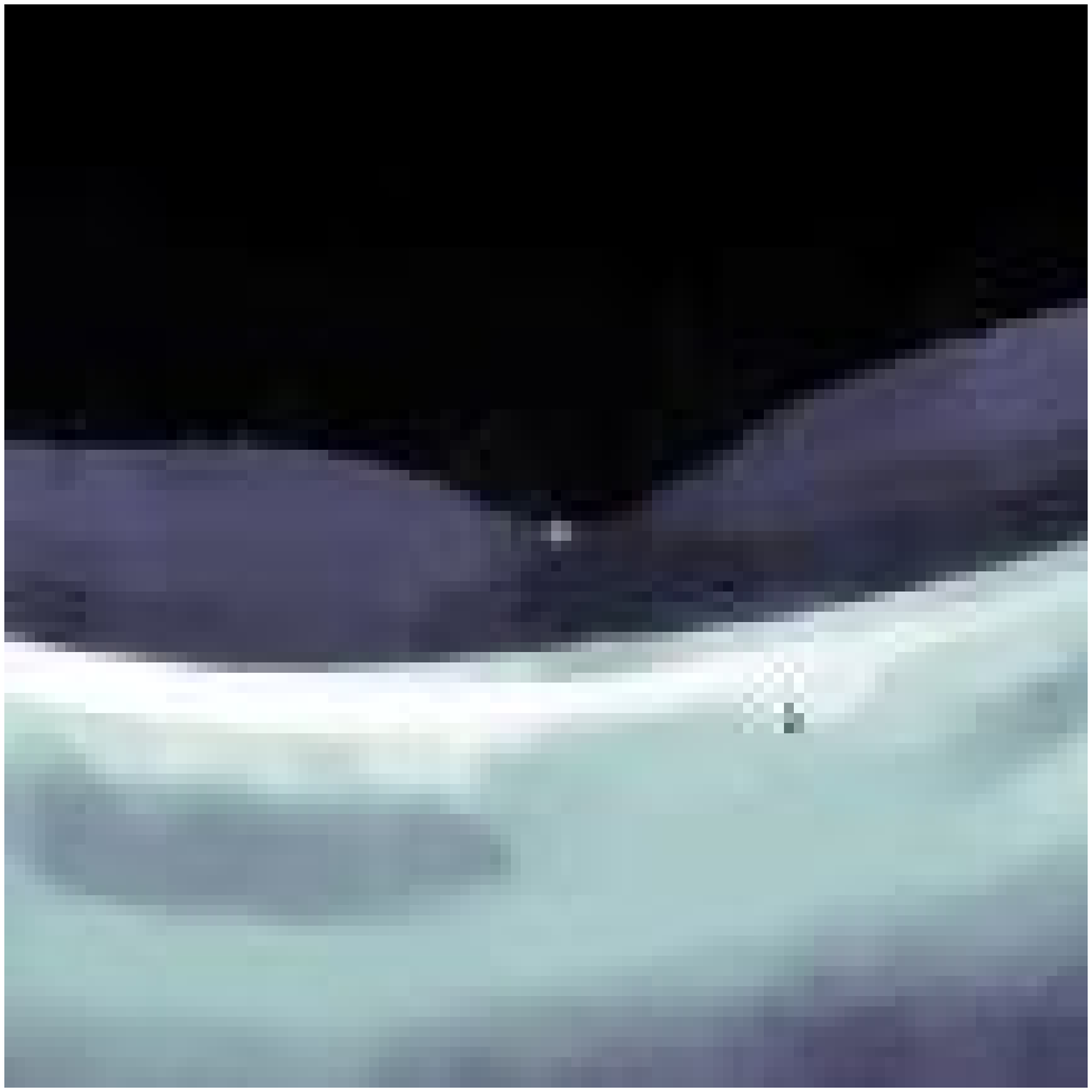}
& \includegraphics[width=50mm, height=50mm]{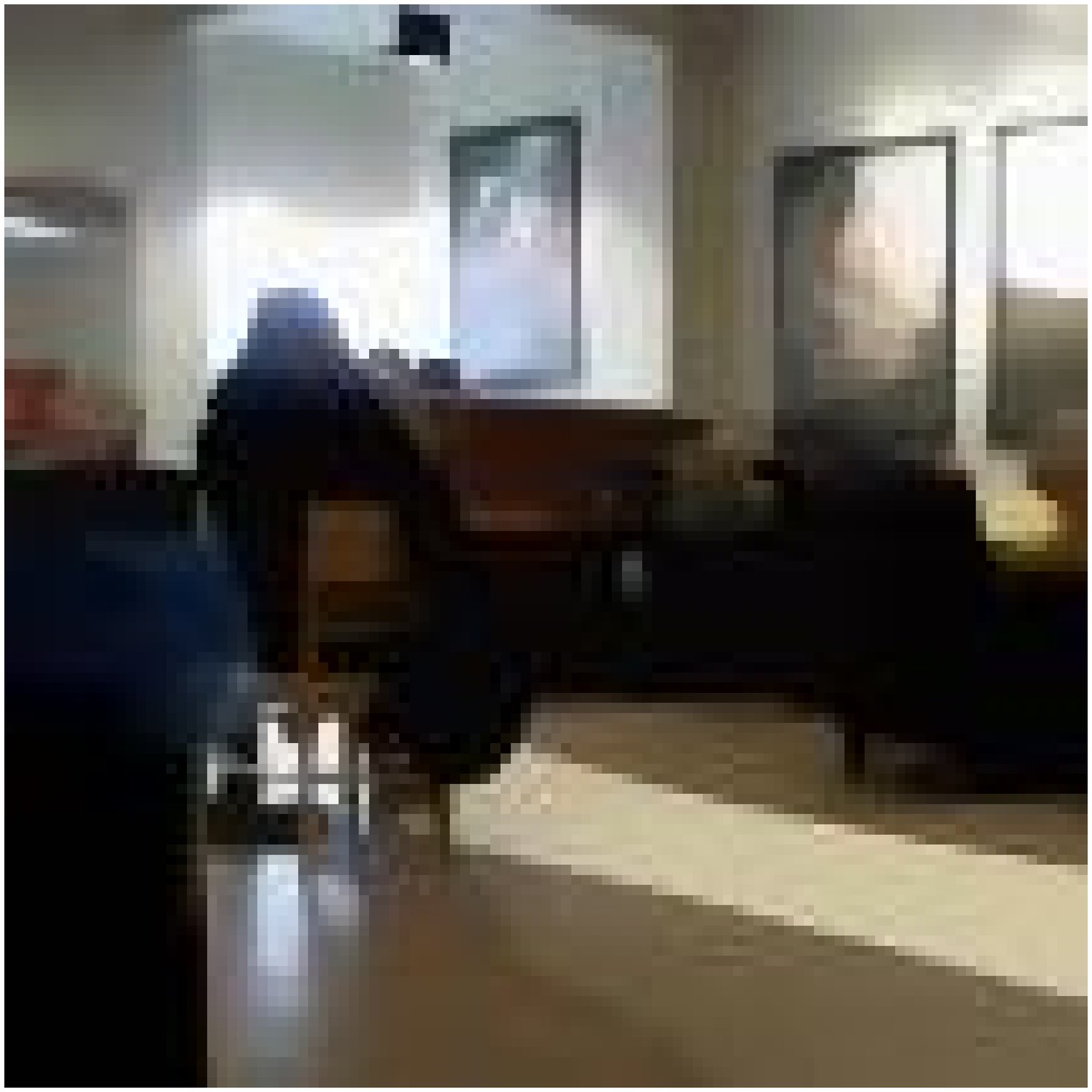}
& \includegraphics[width=50mm, height=50mm]{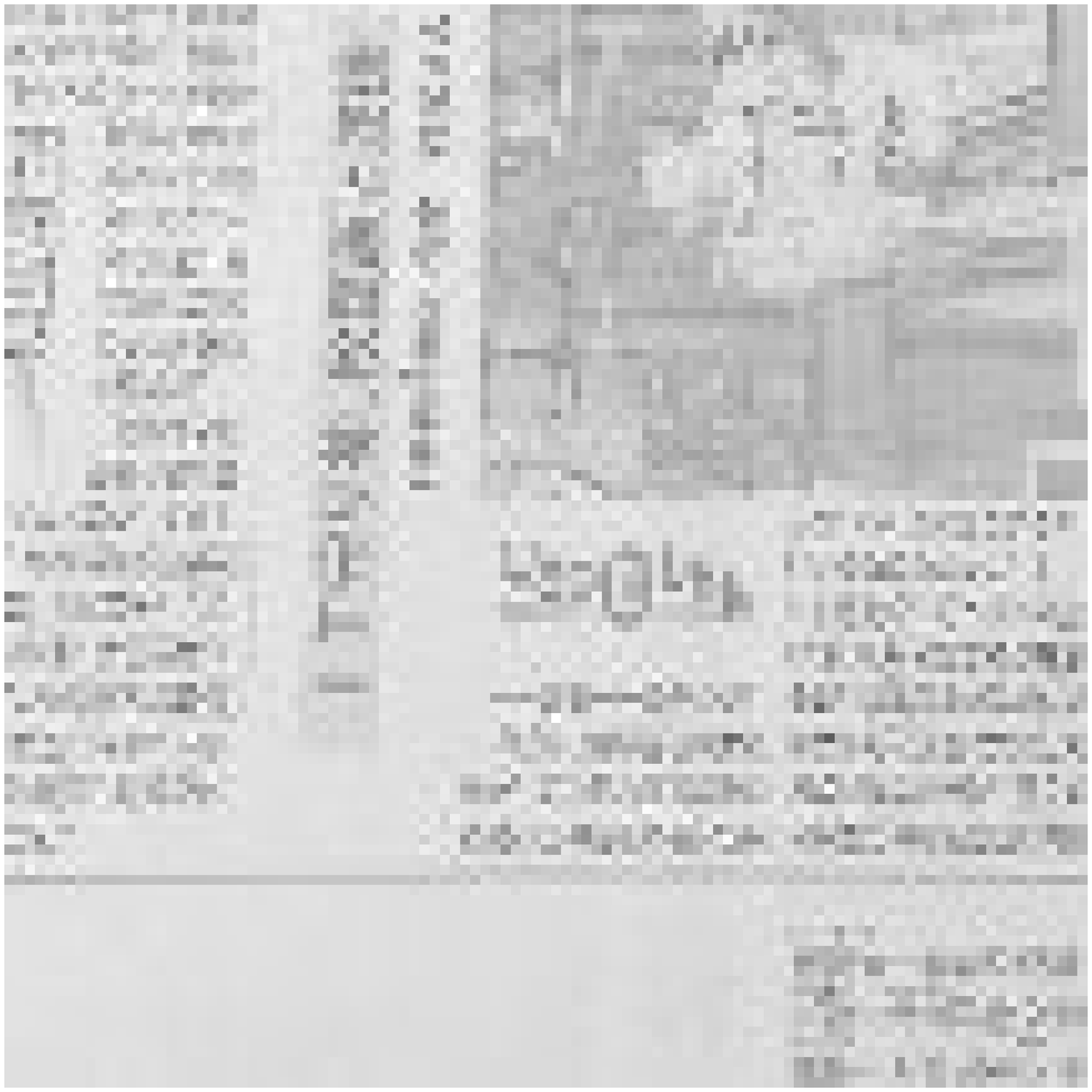}\\
Denoised image that preserved key features.
& Global features
& Enhanced scan
\end{tabular}
\end{center}

\label{fig:real}
\end{figure}

\begin{figure}[H]

\begin{center}
\begin{tabular}{*{3}{C{5.5cm}}}
\includegraphics[width=56mm, height=42mm]{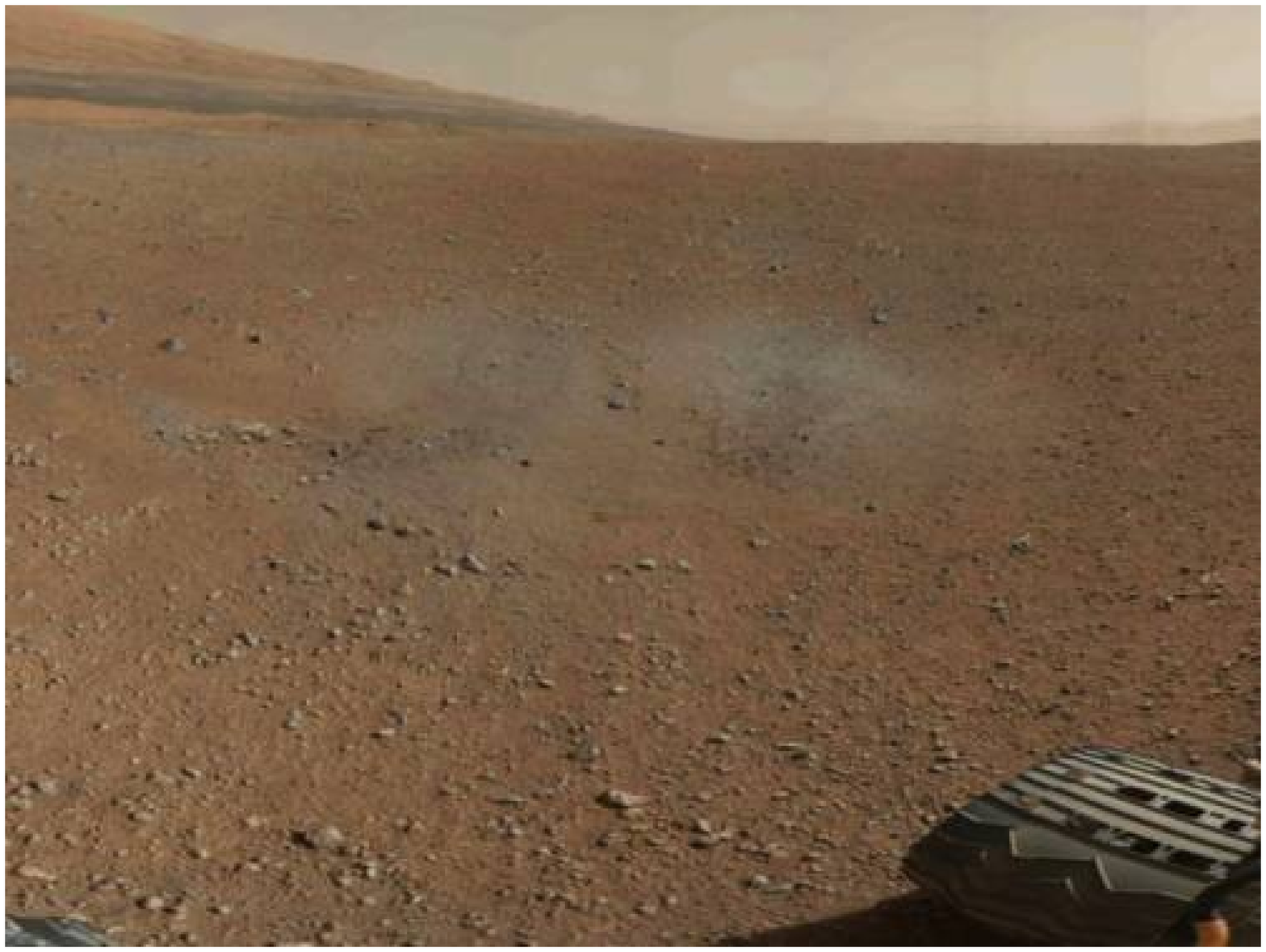} &
\includegraphics[width=42mm, height=28mm]{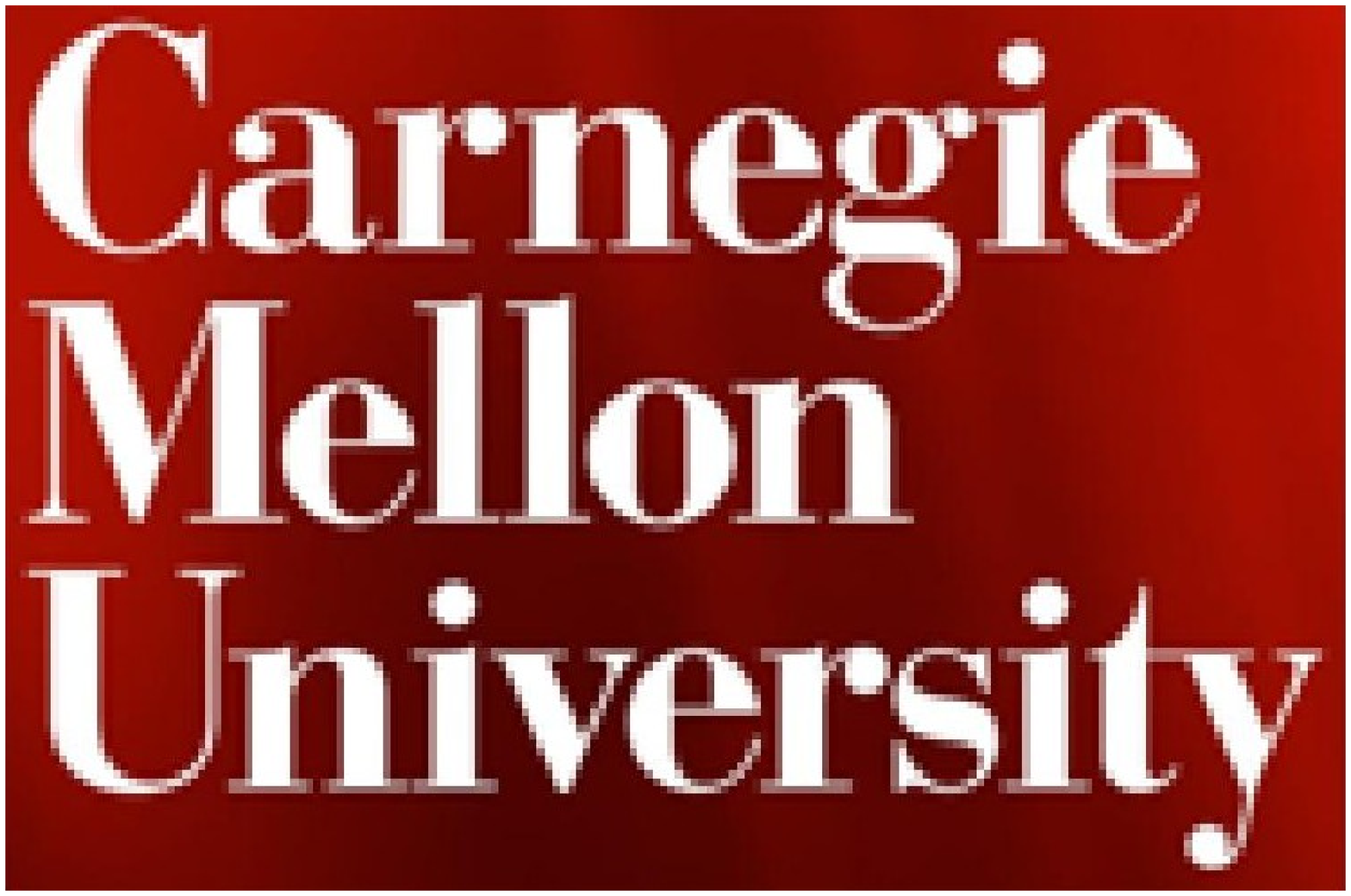} &
\includegraphics[width=56mm, height=42mm]{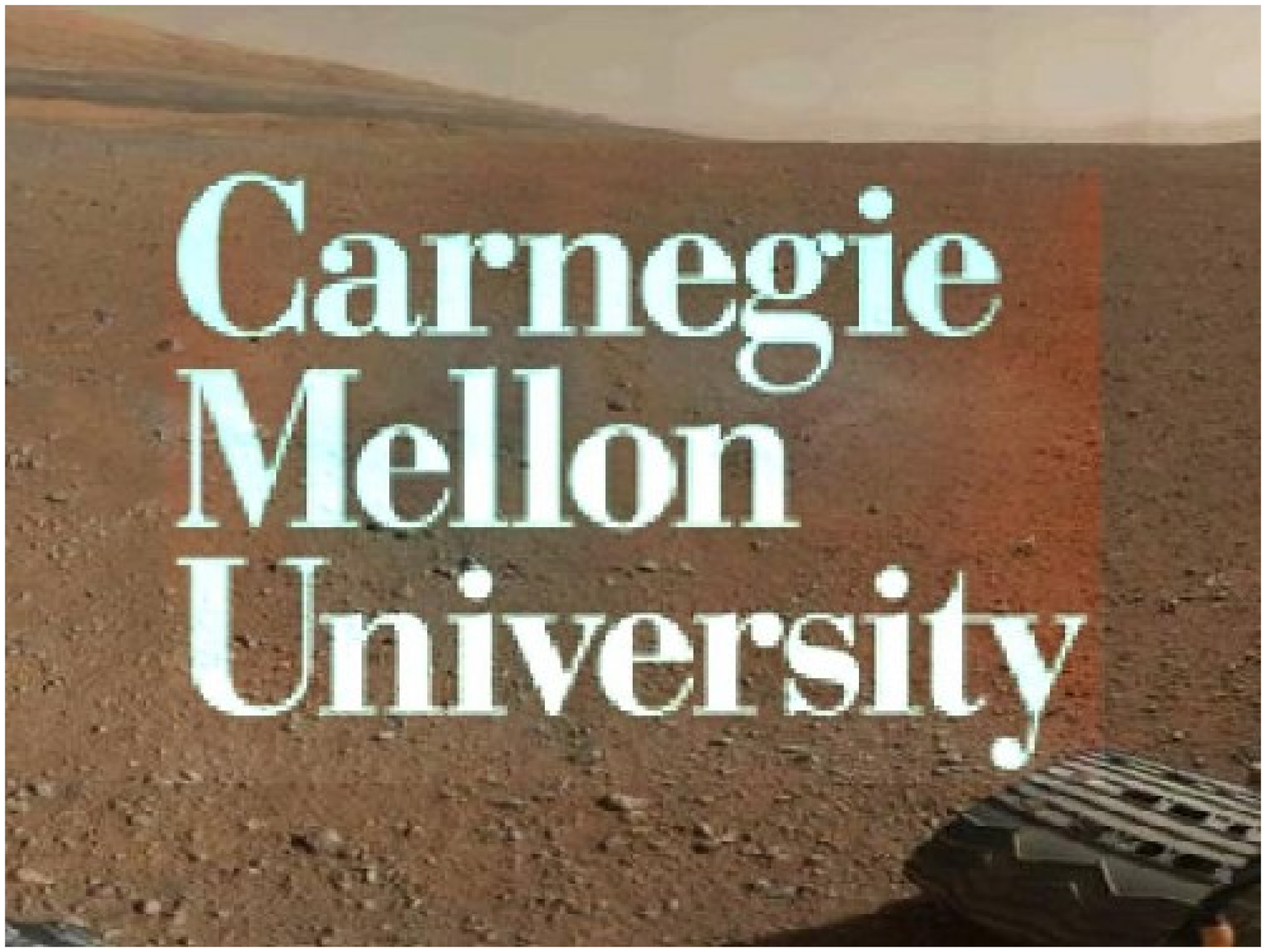} \\
Image of Mars from \textit{Curiosity} &
CMU Logo &
Will it Blend?
\end{tabular}
\end{center}

\label{fig:poisson}
\caption{Example of Poisson Image Editing}
\end{figure}

An example of Poisson Image Editing mentioned in Section
\ref{subsec:poisson} is shown in Figure \ref{fig:poisson}.
The specific application is seamless cloning as described in Section 3
of \cite{Perez03a}, which aims to insert complex objects into another image.
Given two images, they are blended by solving the discrete poisson
equation based on a mix of their gradients and boundary values.
We also added $L_2$ constraints on different parts of the image to
give a smoother result.

\section{Remarks}
\label{sec:conclusion}

We believe that the ability of our algorithm to
encompass many of the current image processing algorithms
represents a major advantage in practice.
It allows the use of a common data structure (the underlying graph)
and subroutine (the SDD solver) for many different tasks in the
image processing pipeline.
Theoretically, this feature is also interesting as it represents
a smooth interpolation between $L_2$ and $L_1$ problems.

The performances of our algorithms depend on $k$, which is the number
of groups in the formulation given in Definition \ref{def:main-prob}.
, it gives them provable runtime guarantees.
Two settings of $k$ are helpful for comparison to previous works.
When $k = 1$, the problem becomes the electrical flow problem,
and the running time of both algorithms are similar to
directly solving the linear system.
This is also the case when there is a small (constant) number of groups.
The other extremum is when each edge belongs to its own group, aka. $k = m$.
Here our approximate algorithm is the same as the minimum $s$-$t$
cut algorithm given in \cite{ChristianoKMST10}, but our analysis for our
almost exact algorithm gives a much worse running time.
This is due to the interior point algorithm generating much more
complicated linear systems, and actually occur when most groups
contain a small number of edges.
As a result, finding faster algorithms with better error guarantees
for problems with intermediate to large values of $k$ is an
interesting direction for future work.
Also, preliminary experimental results such as the ones
from Section \ref{sec:experiments} show that
more aggressive reweightings of edges lead to much faster convergence
than what we showed for our two algorithms.
Therefore a suite of examples where the theoretical guarantees are
tight would also give a better understanding of the interplay
between multiplicative updates and quadratic minimization.

One other consequence of this dependency on $k$ is that
although the problem with smaller number of groups is no longer
captured by linear optimization, the minimum
$s$-$t$ cut problem - that still falls within the framework of linear
optimization is in some sense the hardest problem in this class. 
Therefore we believe that the grouped least squares problem is a
natural interpolation between the $L_1$ and $L_2$ problems, and has
potential as an intermediate subroutine in graph algorithms.

\section*{Acknowledgements}
The authors would like to thank Jerome Darbon, Stanley Osher, Aarti Singh and Ryan Tibshirani for pointing them to works that are relevant to the grouped least squares framework,
and also an anonymous reviewer of a previous submission of this paper for pointing
out an alternate view of the proof of Theorem \ref{thm:algoapprox}.

%\input{weights}

%% \thispagestyle{empty}           % only put this _after_ \maketitle

%% \pagebreak                      % trick to borrow a page
%% \setcounter{page}{1}

% Please use sec:abc thm:abc and lem:abc for all labels.  Each section (and
% subsection), theorem and lemma must have a meaningful label.

%\vfill                          % balance columns
\begin{spacing}{0.7}
  \begin{small}
    \bibliographystyle{alpha}
    \newcommand{\etalchar}[1]{$^{#1}$}

%\bibliography{../../../Bibtex/ref,../../../Bibtex/rpeng,../../../Bibtex/J-koutis,../../../Bibtex/ikoutis,../../../Bibtex/miller,../../../Bibtex/vision,../../../Bibtex/vision2}
  \end{small}
\end{spacing}

\begin{appendix}
\section{Proofs about Graph Problems as Minimizing LASSO Objectives}
\label{sec:graphproofs}

In this section we give formal proofs that show the shortest path problem
is an instance of LASSO and the minimum cut problem is an instance of
fused-LASSO.

It's worth noting that our proofs do not guarantee that the answers
returned are a single path or cut.
In fact, when multiple solutions have the same value it's possible for
our algorithm to return a linear combination of them.
However, we can ensure that the optimum solution is unique 
using the Isolation Lemma of Mulmuley, Vazarani and Vazarani
\cite{MulmuleyVV87} while only incurring polynomial increase in edge lengths/weights.
This analysis is similar to the one in Section 3.5 of \cite{DaitchS08}
for finding a unique minimum cost flow, and is omitted here.

We prove the two claims in Fact \ref{fac:reductionslasso} about
shortest path and minimum cut separately in
Lemmas \ref{lem:shortestpath} and \ref{lem:mincut}.

\begin{lemma} \label{lem:shortestpath}
Given a $s$-$t$ shortest path instance in an undirected graph where
edge lengths $\textbf{l}: E \rightarrow \Re^{+}$ are integers between
$1$ and $n^d$.
There is a LASSO minimization instance where all entries are bounded by
$n^{O(d)}$ such that the value of the LASSO minimizer is within $1$ of
the optimum answer.
\end{lemma}

\Proof
Our reductions rely crucially on the edge-vertex incidence matrix,
which we denote using $\edgevertex$.
Entries of this matrix are defined as follows:

\begin{align}
\edgevertex_{e, u} =
\left\{
\begin{array}{lr}
-1 & \text{if u is the head of e} \\
1 & \text{if u is the tail of e} \\
0 & \text{otherwise}
\end{array}
\right.
\end{align}

We first show the reduction for shortest path.
Then a path from $s$ to $t$ corresponds to a flow value assigned to
all edges, $\flowv: E \rightarrow \Re$ such that
$\edgevertex^T \flowv = \chi_{s,t}$.
If we have another flow $\flowv'$ corresponding to any path from $s$
to $t$, then this constraint can be written as:

\begin{align}
||\edgevertex^T \flowv - \chi_{s,t}||_2 = & \zerov \\
||\edgevertex^T (\flowv - \flowv')||_2 = & \zerov \nonumber \\
||\flowv - \flowv'||_{\edgevertex \edgevertex^T} = & \zerov \\
\end{align}

The first constraint is closer to the classical LASSO problem while
the last one is within our definition of grouped least squares problem.
The length of the path can then be written as $\sum_{e} l_e |f_e|$.
Weighting these two terms together gives:

\begin{align}
\min_{\flowv} \lambda ||\flowv - \flowv'||^2_{\edgevertex \edgevertex^T}
+ \sum_{e} l_e |f_e|
\end{align}

Where the maximum entry is bounded by $\max \{ n^d, n^2 \lambda \}$.
Clearly its objective is less than the length of the shortest path, let this
solution be $\optflowv$.
Then since the total objective is at most $n^{d + 1}$, we have that
the maximum deviation between $\edgevertex^T \optflowv$
and $\chi_{s,t}$ is at most $n^{d + 1} / \lambda$.
Then given a spanning tree, each of these deviations can be routed
to $s$ or $t$ at a cost of at most $n^{d + 1}$ per unit of flow.
Therefore we can obtain $\flowv'$ such that $\edgevertex^T \flowv' = \chi_{s,t}$
whose objective is bigger by at most $n^{2d + 2} / \lambda$.
Therefore setting $\lambda = n^{2d+2}$ guarantees that our objective
is within $1$ of the length of the shortest path, while the maximum
entry in the problem is bounded by $n^{O(d)}$.
\QED

We now turn our attention to the minimum cut problem, which can
be formulated as finding a vertex labeling $\vecx^{(vert)}$
where $\vecx^{(vert)}_s = 0$, $\vecx^{(vert)}_t = 1$
and the size of the cut,
$\sum_{uv \in E} |\vecx^{(vert)}_u - \vecx^{(vert)}_v|$.
Since the $L_1$ term in the objective can incorporate single variables,
we use an additional vector $\vecx^{(edge)}$
to indicate differences along edges.
The minimum cut problem then becomes minimizing
$|\vecx^{(edge)}|$ subject to the constraint that
$\vecx^{(edge)} = \edgevertex' \vecx^{(vert)'} + \edgevertex \chi_t$,
where $\edgevertex'$ and $\vecx^{(vert)'}$ are restricted to vertices
other than $s$ and $t$ and $\chi_t$ is the indicator vector that's $1$ on $t$.
The equality constraints can be handled similar to the shortest
path problem by increasing the weights on the first term.
One other issue is that $|\vecx^{(vert)'}|$ also appears in the objective
term, and we handle this by scaling down $\vecx^{(vert)'}$, or equivalently
scaling up $\edgevertex'$.

\begin{lemma} \label{lem:mincut}
Given a $s$-$t$ minimum cut instance in an undirected graph where
edge weights $\textbf{w}: E \rightarrow \Re^{+}$ are integers between
$1$ and $n^d$.
There is a LASSO minimization instance where all entries are bounded by
$n^{O(d)}$ such that the value of the LASSO minimizer is within $1$ of
the minimum cut.
\end{lemma}

\Proof

Consider the following objective, where $\lambda_1$
and $\lambda_2$ are set to $n^{d + 3}$ and $n^2$:

\begin{align}
\lambda_1||\lambda_2 \edgevertex' \vecx^{(vert)'} + \edgevertex \chi_t - \vecx^{(edge)} ||^2_2
+ |\vecx^{(vert')}|_1 + |\vecx^{(edge)}|_1
\end{align}

Let $\bar{\vecx}$ be an optimum vertex labelling, then setting $\vecx^{(vert')}$
to the restriction of $n^{-2} \bar{\vecx}$ on vertices other than $s$ and $t$
and $\vecx^{(edge)}$ to $\edgevertex \vecx^{(vert)}$ makes the first term $0$.
Since each entry of $\bar{\vecx}$ is between $0$ and $1$,
the additive increase caused by $|\vecx^{(vert')}|_1$ is at most $1/n$.
Therefore this objective's optimum is at most $1/n$ more than the size of
the minimum cut.

For the other direction, consider any solution $\vecx^{(vert)'}, \vecx^{(edge)'}$
whose objective is at most $1/n$ more than the size of the minimum cut.
Since the edge weights are at most $n^{d}$ and $s$ has degree at most $n$,
the total objective is at most $n^{d+1} + 1$.
This gives:
\begin{align}
||\lambda_2 \edgevertex' \vecx^{(vert)'} + \edgevertex \chi_t - \vecx^{(edge)} ||^2_2
\leq & n^{-1} \\
||\lambda_2 \edgevertex' \vecx^{(vert)'} + \edgevertex \chi_t - \vecx^{(edge)} ||_1
\leq & ||\lambda_2 \edgevertex' \vecx^{(vert)'} + \edgevertex \chi_t - \vecx^{(edge)} ||_2
\leq n^{-1/2}
\end{align}

Therefore changing $\vecx^{(edge)}$ to
$\lambda_2 \edgevertex' \vecx^{(vert)'} + \edgevertex \chi_t$
increases the objective by at most $n^{-1/2} < 1$.
This gives a cut with weight within $1$ of the objective value and
completes the proof.
\QED

We can also show a more direct connection between the minimum cut
problem and the fused LASSO objective,
where each absolute value term may contain a linear combination of variables.
This formulation is closer to the total variation objective, and is also
an instance of the problem formulated in Definition \ref{def:main-prob}
with each edge in a group.

\begin{lemma}
\label{lem:mincutfused}
The minimum cut problem in undirected graphs can be written as
an instance of the fused LASSO objective.
\end{lemma}

\Proof
Given a graph $G = (V, E)$ and edge weights $\costv$,
the problem can be formulated as labeling the vertices of $V \setminus \{s, t\}$
in order to minimize:

\begin{align}
\sum_{uv \in E, u, v \in V \setminus \{s, t\}} \cost_{uv}|x_u - x_v|
+ \sum_{su \in E, u \in V \setminus \{s, t\}} \cost_{su} |x_u - 0|
+ \sum_{ut \in E, u \in V \setminus \{s, t\}} \cost_{ut} |x_u - 1|
\end{align}

\QED

\section{Multiplicative Weights Based Approximate Algorithm}
\label{sec:multiplicativeweights}

In this section we show that the approximate algorithm described in
Section \ref{subsec:approx} finds a solution close to the optimum.
We first show that if $\lambda^{(t)}$ as defined on Line \ref{ln:lambda}
of Algorithm \ref{alg:algoapprox} is an upper bound for
$\linearprimal(\vecx^{(t)}$.
This is crucial in our use of it as the normalizing factor in our update
step on Line \ref{ln:update}.

\begin{lemma}
\label{lem:lambda}
In all iterations we have:

\begin{align*}
\linearprimal(\vecx^{(t)}) \leq \lambda^{(t)}
\end{align*}
\end{lemma}

\Proof
By the Cauchy-Schwarz inequality we have:

\begin{align}
(\lambda^{(t)})^2
= & \left( \sum_i \weight_i^{(t - 1)} \right)
 \left( \sum_i \frac{1}{\weight_i^{(t - 1)}} || \vecx^{(t)} - \fixedv_i ||^2 _{\group_i} \right)
\nonumber \\
\geq & \left( \sum_i  || \vecx^{(t)} - \fixedv_i ||_{\group_i} \right)^2
\nonumber \\
= & \linearprimal(\vecx^{(t)})^2
\end{align}

Taking square roots of both sides completes the proof.
\QED

At a high level, the algorithm assigns weights $\weight_i$ for each group,
and iteratively reweighs them for $N$ iterations.
Recall that our key potential functions are $\mu^{(t)}$ which is the sum of
weights of all groups, and:

\begin{align}
\nu^{(t)} = \frac{1}{\opt} \sum_i || \optvecx - \fixedv_i ||_{\group_i}
        \log(\weight_i^{(t)}) \label{eq:nudef}
\end{align}

Where $\optvecx$ is a solution such that $\linearprimal(\optvecx) = \opt$.
We will show that if $\linearprimal(\vecx^{(t)})$, or in turn
$\lambda^{(t)}$ is large, then $\nu^{(t)}$ increases at a rate
substantially faster than $\mu^{(t)}$.
These bounds, and the relation between $\mu^{(t)}$ and $\nu^{(t)}$ is
summarized below:

\begin{lemma}
\label{lem:potentials}
\begin{enumerate}
\item
\label{part:lowervsupper}
\begin{align}
    \nu^{(t)} \leq \log(\mu^{(t)}) \label{eq:numu}
\end{align}
\item
\label{part:upperincrease}
\begin{align}
    \mu^{(t)} \leq \left( 1 + \frac{\epsilon(1+ 2\epsilon) }{\rho} t \right) \mu^{(t-1)} \label{eq:muupper}
\end{align}
and
\begin{align}
    \log(\mu^{(t)}) \leq \frac{\epsilon(1 + 2\epsilon) }{\rho} t+ \log{k} \label{eq:muupperlog}
\end{align}
\item
\label{part:lowerincrease}
If in iteration $t$, $\lambda^{(t)} \geq (1 + 10 \epsilon) \opt$
and
$||\vecx - \fixedv_i||_{\group_i} \leq
    \rho \frac{ \weight_i^{(t - 1)} }{\mu^{(t-1)}} \lambda^{(t)}$
for all groups $i$, then:

\begin{align}
    \nu^t \geq \nu^{(t-1)} + \frac{\epsilon(1 + 9 \epsilon)}{\rho} \label{eq:nulower}
\end{align}
\end{enumerate}
\end{lemma}

The relationship between the upper and lower potentials can be established
using the fact that $\weight_i$ is non-negative:

\Proofof{Lemma \ref{lem:potentials}, Part \ref{part:lowervsupper}}
\begin{align}
\nu^{(t)}
= & \frac{1}{\opt} \sum_i
    ||\optvecx - \fixedv_i||_{\group_i} \log(\weight_i^{(t)}) \nonumber \\
\leq & \frac{1}{\opt} \sum_i ||\optvecx - \fixedv_i||_{\group_i}
    \log\left( \sum_j \weight_j^{(t)} \right) \nonumber \\
= & \log(\mu^{(t)}) \left( \frac{1}{\opt} \sum_i
    ||\optvecx - \fixedv_i||_{\group_i}\right) \nonumber \\
\leq & \log(\mu^{(t)})
\end{align}
\QED

Part \ref{part:upperincrease} follows directly from the local behavior of
the $\log$ function:

\Proofof{Lemma \ref{lem:potentials}, Part \ref{part:upperincrease}}
The update rules gives:

\begin{align}
\mu^{(t)}
= & \sum_{i} \weight^{(t)}_i \nonumber \\
= & \sum_{i} \weight^{(t-1)}_i + \left( \frac{\epsilon}{\rho} \frac{||\vecx^{(t)} - \fixedv_i||_{\group_i}}{\lambda^{(t)}} + \frac{2\epsilon^2}{k\rho} \right) \mu^{(t-1)} 
\nonumber \\ & \qquad \mbox{by update rule on Line \ref{ln:update} of \textsc{GroupedLeastSquares}} \nonumber \\
= & \mu^{(t - 1)} + \frac{\epsilon}{\rho}
    \frac{\sum_i ||\vecx^{(t)} - \fixedv_i||_{\group_i}}{\lambda^{(t)}} \mu^{(t - 1)}
    + \sum_{i} \frac{2\epsilon^2}{k \rho} \mu^{(t - 1)}
 \nonumber \\
= & \mu^{(t - 1)} + \frac{\epsilon}{\rho}
    \frac{\linearprimal(\vecx^{(t)})}{\lambda^{(t)}} \mu^{(t - 1)}
    + \frac{2\epsilon^2}{\rho} \mu^{(t - 1)}
 \nonumber \\
\leq & \mu^{(t - 1)} + \frac{\epsilon}{\rho} \mu^{(t - 1)}
    + \frac{2\epsilon^2}{\rho} \mu^{(t - 1)}
 \qquad  \mbox{By Lemma \ref{lem:lambda}} \nonumber \\
= & \left( 1 + \frac{\epsilon(1 + 2\epsilon)}{\rho}\right) \mu^{(t-1)}
\end{align}

Using the fact that $1+x \leq \exp(x)$ when $x \geq 0$ we get:

\begin{align*}
\mu^{(t)}
\leq & \exp \left( \frac{\epsilon(1 + 2\epsilon)}{\rho} \right) \mu^{(t-1)} \nonumber \\
\leq & \exp \left( t \frac{\epsilon(1 + 2\epsilon)}{\rho} \right) \mu^{(0)} \nonumber \\
= & \exp \left( t \frac{\epsilon(1 + 2\epsilon)}{\rho} \right) k
\end{align*}

Taking logs of both sides gives Equation \ref{eq:muupperlog}.

\QED

This upper bound on the value of $\mu^t$ also allows us to show that the balancing
rule keeps the $w^t_i$s reasonably balanced within a factor of $k$ of each other.
The following corollary can also be obtained.

\begin{corollary}
\label{cor:balance}
The weights at iteration $t$ satisfy $\weight^{(t)}_i \geq \frac{\epsilon}{k} \mu^{(t)}$.
\end{corollary}

\Proof

The proof is by induction on $t$.
When $t = 0$ we have $\weight^{(0)}_i = 1$,
$\mu^{(0)} = k$ and the claim follows from $\frac{\epsilon}{k} k = \epsilon < 1$.
When $t > 1$, we have:
\begin{align}
\weight^{(t)}_i
\geq & \weight^{(t-1)}_i + \frac{2\epsilon^2}{k\rho} \mu^{(t-1)}
\qquad \mbox{By line \ref{ln:update}} \nonumber \\
\geq & \left( \frac{\epsilon}{k} + \frac{2\epsilon^2}{k\rho} \right) \mu^{(t-1)}
\qquad \mbox{By the inductive hypothesis} \nonumber \\
= & \frac{\epsilon}{k} \left( 1 + \frac{2\epsilon}{\rho} \right) \mu^{(t-1)} \nonumber \\
\geq & \frac{\epsilon}{k}
    \left( 1 + \frac{\epsilon(1 + 2\epsilon)}{\rho} \right) \mu^{(t-1)} \nonumber \\
\geq & \frac{\epsilon}{k} \mu^{(t)}
\qquad \mbox{By Lemma \ref{lem:potentials}, Part \ref{part:upperincrease}}
\end{align}
\QED

The proof of Part \ref{part:lowerincrease} is the key part of our analysis.
The first order change of $\nu^{(t)}$ is written as a sum of products of
$L_2$ norms, which we analyze via. the fact that $\vecx^{(t)}$ is the
solution of a linear system from the quadratic minimization problem.

\Proofof{Lemma \ref{lem:potentials}, Part \ref{part:lowerincrease}}

We make use of the following known fact about the behavior of the log function
around $1$:

\begin{fact}
\label{fact:log}
If $0 \leq x \leq \epsilon$, then $\log(1+x) \geq (1-\epsilon)x$.
\end{fact}

\begin{align}
\nu^{(t)} - \nu^{(t-1)}
 = & \frac{1}{\opt} \sum_{1 \leq i \leq k} || \optvecx - \fixedv_i ||_{\group_i} \log \left( \weight_i^{(t)} \right)
    - \frac{1}{\opt} \sum_{1 \leq i \leq k} || \optvecx - \fixedv_i ||_{\group_i} \log \left( \weight_i^{(t-1)} \right)
\qquad \mbox{By Equation \ref{eq:nudef} } \nonumber \\
= & \frac{1}{\opt} \sum_{1 \leq i \leq k}
    || \optvecx - \fixedv_i ||_{\group_i}
        \log \left( \frac{\weight_i^{(t)}}{\weight_i^{(t - 1)}} \right) \nonumber \\
\geq & \frac{1}{\opt} \sum_{1\leq i \leq k}
    || \optvecx - \fixedv_i ||_{\group_i}
        \log \left( 1 + \frac{\epsilon}{\rho}
                \frac{||\vecx^{(t)} - \fixedv_i||_{\group_i}}{\lambda^{(t)}}
                      \frac{\mu^{(t - 1)} }{\weight^{(t - 1)}_i} \right)
\qquad \mbox{By update rule in line \ref{ln:update}} \nonumber \\
\geq & \frac{1}{\opt} \sum_{1 \leq i \leq k}
    || \optvecx - \fixedv_i ||_{\group_i}
        \frac{\epsilon (1 - \epsilon)}{\rho}
                \frac{||\vecx^{(t)} - \fixedv_i||_{\group_i}}{\lambda^{(t)}}
                      \frac{\mu^{(t - 1)} }{\weight^{(t - 1)}_i}
\qquad \mbox{Since $\log(1+x) \geq (1-\epsilon)x$ when $0 \leq x \leq \epsilon$}
\nonumber \\
= & \frac{\epsilon (1 - \epsilon) \mu^{(t - 1)}_i }{\rho \opt \lambda^{(t)}}
    \sum_{1 \leq i \leq k} \frac{1}{\weight^{(t - 1)}_i}
          || \optvecx - \fixedv_i ||_{\group_i} ||\vecx^{(t)} - \fixedv_i||_{\group_i}
\label{eq:lowerchangelinear}
\end{align}

Since $\group_i$ forms a P.S.D norm, by the Cauchy-Schwarz inequality we have:

\begin{align}
|| \optvecx - \fixedv_i ||_{\group_i} ||\vecx^{(t)} - \fixedv_i||_{\group_i}
\geq & (\optvecx - \fixedv_i)^T \group_i (\vecx^{(t)} - \fixedv_i) \\
= & ||\vecx^{(t)} - \fixedv_i||_{\group_i}^2 +
       (\optvecx - \vecx)^T \group_i (\vecx^{(t)} - \fixedv_i)
\end{align}

Recall from Lemma \ref{lem:quadminimization} that since $\vecx^{(t)}$
is the minimizer to $\quadprimal(\weightv^{(t - 1)})$, we have

\begin{align}
\left( \sum_{i} \weightv^{(t - 1)}_i \group_i \right) \vecx^{(t)}
     = & \sum_{i} \weightv^{(t - 1)} \fixedv_i \\
\left( \sum_{i} \weightv^{(t - 1)}_i \group_i \right) ( \vecx^{(t)} - \fixedv_i )
     = & \zerov\\
(\optvecx - \vecx^{(t)})^T
    \left( \sum_{i} \weightv^{(t - 1)}_i \group_i \right) ( \vecx^{(t)} - \fixedv_i )
     = & 0\\
\end{align}

Substituting this into Equation \ref{eq:lowerchangelinear} gives:

\begin{align}
\nu^{(t)} - \nu^{(t-1)}
\geq & \frac{\epsilon (1 - \epsilon) \mu^{(t - 1)}_i }{\rho \opt \lambda^{(t)}}
    \sum_{i} \frac{1}{\weight_i^{(t - 1)}} ||\vecx^{(t)} - \fixedv_i||_{\group_i}^2
 \nonumber \\
= & \frac{\epsilon (1 - \epsilon) }{\rho \opt \lambda^{(t)}}
    \mu^{(t - 1)} \quadprimal(\weightv, \vecx^{(t)}) \nonumber \\
= & \frac{\epsilon (1 - \epsilon) }{\rho \opt \lambda^{(t)}} (\lambda^{(t)})^2
	\qquad \mbox{By definition of $\lambda^{(t)}$ on Line \ref{ln:lambda}} \nonumber \\
\geq & \frac{\epsilon (1 - \epsilon) (1 + 10 \epsilon) }{\rho}
    \qquad \mbox{By assumption that $\lambda^{(t)}> (1 + 10 \epsilon )\opt$} \nonumber \\
\geq & \frac{\epsilon (1 + 8 \epsilon)}{\rho}
\end{align}

Since the iteration count largely depends on $\rho$, it suffices to provide bounds
for $\rho$ over all the iterations.
The proof makes use of the following lemma about the properties of electrical flows,
which describes the behavior of modifying the weights of a group $S_i$ that has
a large contribution to the total energy.
It can be viewed as a multiple-edge version of Lemma 2.6 of \cite{ChristianoKMST10}.

\begin{lemma}
\label{lem:quadincrease}
Assume that $\epsilon^2 \rho^2 < 1/10 k$ and $\epsilon < 0.01$ and
let $\vecx^{(t - 1)}$ be the minimizer for $\quadprimal(\weightv^{(t - 1)})$.
Suppose there is a group $i$ such that
$||\vecx^{(t - 1)} - \fixedv_i ||_{\group_i}
\geq \rho \frac{ \weight_i^{(t - 1)} }{\mu^{(t-1)}} \lambda^{(t)}$, then 
\begin{align*}
\optquad(\weightv^{(t)}) \leq
\exp \left(-\frac{\epsilon^2 \rho^2}{2k}\right)  \optquad(\weightv^{(t)})
\end{align*}
\end{lemma}

\Proof

We first show that group $i$ contributes a significant portion to
$\quadprimal(\weightv^{(t - 1)}, \vecx^{(t - 1)})$.
Squaring both sides of the given condition gives:

\begin{align}
||\vecx^{(t - 1)} - \fixedv_i ||^2_{\group_i}
\geq & \rho^2 \frac{ (\weight_i^{(t - 1)})^2 }{(\mu^{(t-1)})^2} (\lambda^{(t)})^2
\nonumber \\
= & \rho^2 \frac{ (\weight_i^{(t - 1)})^2 }{(\mu^{(t-1)})^2} \mu^{(t - 1)} 
  \quadprimal(\weightv^{(t - 1)}, \vecx^{(t - 1)}) \\
\frac{1}{ \weight_i^{(t - 1)} } ||\vecx^{(t - 1)} - \fixedv_i ||_{\group_i}
\geq & \rho ^2  \frac{ \weight_i^{(t - 1)}}{ \mu^{(t-1)}} 
  \quadprimal(\weightv^{(t - 1)}, \vecx^{(t - 1)}) \nonumber \\
\geq & \frac{\epsilon \rho^2}{k} \quadprimal(\weightv^{(t - 1)}, \vecx^{(t - 1)})
\qquad \mbox{By Corollary \ref{cor:balance}}
\end{align}

Also, by the update rule we
have $\weight^{(t)}_i \geq (1 + \epsilon) \weight^{(t - 1)}_i$ and
$\weight^{(t)}_j \geq \weight^{(t - 1)}_j$ for all $1 \leq j \leq k$.
So we have:

\begin{align}
\optquad(\weightv^{(t)})
\leq & \quadprimal(\weightv^{(t)}, \vecx^{(t - 1)}) \nonumber \\
= & \quadprimal(\weightv^{(t)}, \vecx^{(t - 1)})
- (1 - \frac{1}{1 + \epsilon}) ||\vecx^{(t - 1)} - \fixedv_i ||^2_{\group_i} \nonumber \\
\leq & 
\quadprimal(\weightv^{(t)}, \vecx^{(t - 1)})
- \frac{\epsilon}{2} ||\vecx^{(t - 1)} - \fixedv_i ||^2_{\group_i} \nonumber \\
\leq &\quadprimal(\weightv^{(t)}, \vecx^{(t - 1)})
- \frac{\epsilon^2 \rho^2}{2k}
   \quadprimal(\weightv^{(t - 1)}, \vecx^{(t - 1)}) \nonumber \\
\leq &\exp \left( -\frac{\epsilon^2 \rho^2}{2k} \right) 
   \quadprimal(\weightv^{(t - 1)}, \vecx^{(t - 1)})
\end{align}

\QED

This means the value of the quadratic minimization problem
can be used as a second potential function.
We first show that it's monotonic and establish rough bounds for it.

\begin{lemma}
\label{lem:roughbounds}
$\optquad(\weightv^{(0)}) \leq n^{3d}$
and $\optquad(\weightv^{(t)})$ is monotonically decreasing in $t$.
\end{lemma}

\Proof
By the assumption that the input is polynomially bounded we have
that all entries of $\fixedv$ are at most $n^d$ and
$\group_i \preceq n^{d} \mident$.
Setting $\vecx_u = 0$ gives $||\vecx - \fixedv_i||_2 \leq n^{d+1}$.
Combining this with the spectrum bound then gives
$||\vecx - \fixedv_i||_{\group_i} \leq n^{2d+1}$.
Summing over all the groups gives the upper bound.

The monotonicity of $\optquad(\weightv^{(t)})$ follows from the
fact that all weights are decreasing.
\QED

Combining this with the fact that $\optquad(\weight^{(N)})$ is not low
enough for termination gives our bound on the total iteration count.

\Proofof{Theorem \ref{thm:algoapprox}}
The proof is by contradiction. Suppose otherwise, since
$\linearprimal(\vecx^{(N)}) > \epsilon$ we have:

\begin{align}
\lambda^{(N)} \geq & (1 + 10 \epsilon) n^{-d} \nonumber \\
\geq & 2 n^{-d} \opt \\
\sqrt{\mu^{(N)} \optquad(\weightv^{(t)})}
\geq & 2 n^{-d} \\
\optquad(\weightv^{(t)})
\geq \frac{4}{n^{-2d} \mu^{(N)}} 
\end{align}

Which combined with $ \optquad(\weightv^{(0)})  \leq n^{3d}$
from Lemma \ref{lem:roughbounds} gives:

\begin{align}
\frac{\optquad(\weightv^{(0)})}{\optquad(\weightv^{(N)}) }
\leq & n^{5d} \mu^{(N)}
\end{align}

By Lemma \ref{lem:potentials} Part \ref{part:upperincrease},
we have:

\begin{align}
\log(\mu^{(N)})
\leq & \frac{\epsilon(1 + \epsilon)}{\rho} N + \log{k} \nonumber \\
\leq & \frac{\epsilon(1 + \epsilon)}{\rho} 10 d \rho \log{n} \epsilon^{-2} + \log{n}
\qquad \mbox{By choice of $N = 10d \rho \log{n} \epsilon^{-2}$}\nonumber \\
= & 10(1 + \epsilon) \epsilon^{-1} \log{n} + \log{n} \nonumber \\
\leq & 10d (1 + 2 \epsilon) \epsilon^{-1} \log{n}
\qquad \mbox{when $\epsilon < 0.01$}
 \end{align}

Combining with Lemma \ref{lem:quadincrease} implies that the number
of iterations where $||\vecx^{(t - 1)} - \fixedv_i ||_{\group_i}
\geq \rho \frac{ \weight_i^{(t - 1)} }{\mu^{(t-1)}} \lambda^{(t)}$
for $i$ is at most:

\begin{align}
\log\left(\mu^{(N)} n^{5d}\right) / \left(\frac{\epsilon^2 \rho^2}{2k}\right)
%\nonumber \\
= & 10d (1 + 3 \epsilon) \epsilon^{-1} \log{n}
    / \left(\frac{2\epsilon^{2/3} }{ k^{1/3}}\right)
\qquad \mbox{By choice of $\rho = 2 k^{1/3} \epsilon^{-2/3}$} \nonumber \\
= & 8 d\epsilon^{-5/3} k^{1/3} \log{n} \nonumber \\
= & 4 d \epsilon^{-1} \rho \log{n}
\leq \epsilon N
\end{align}

This means that we have
$||\vecx^{(t - 1)} - \fixedv_i ||_{\group_i}
\leq \rho \frac{ \weight_i^{(t - 1)} }{\mu^{(t-1)}} \lambda^{(t)}$
for all $1 \leq i \leq k$ for at least $(1-\epsilon)N$ iterations
and therefore by Lemma \ref{lem:potentials} Part \ref{part:lowerincrease}:

\begin{align}
\nu^{(N)}
\geq & \nu^{(0)} + \frac{\epsilon(1 + 8 \epsilon)}{\rho}(1-\epsilon)N
> \mu^{(N)}
\end{align}

Giving a contradiction.
\QED

\section{Solving Linear Systems from Interior Point Algorithms}

\label{sec:hessian}

In this section we take a closer look at the linear system solves required by
Lemma \ref{lem:interiorpoint}, specifically the Hessians of the objective
given in Equation \ref{eq:logbarrier}.
We show that for small values $k$, having access to an efficient subroutines
for solving the quadratic minimization problems gives improvements
over general interior-point algorithms.

\Proofof{Theorem \ref{thm:hessiansolve}}

We first consider the barrier function corresponding to each group, $\phi(\vecx, y_i)$.
Its gradient is:

\begin{align}
\nabla \phi(\vecx, y_i)
= & \nabla - \log \left( y_i^2 -
     (\vecx - \fixedv_i)^T\group_i (\vecx - \fixedv_i) \right) \nonumber \\
= & \frac{2}{y_i^2 - ||\vecx - \fixedv_i||_{\group_i}^2 }
\left[
\begin{array}{c} 
\group_i (\vecx - \fixedv_i) \\
-y_i
\end{array}
\right]
\end{align}

and its Hessian is:

\begin{align}
\nabla^2 \phi(\vecx, y_i)
= & \frac{2}{(y_i^2 - ||\vecx - \fixedv_i||_{\group_i}^2)^2}
\left[
\begin{array}{cc} 
\group_i (y_i^2 - ||\vecx - \fixedv_i||_{\group_i}^2)
+ 2\group_i (\vecx - \fixedv_i) (\vecx - \fixedv_i)^T \group_i & 
2 y_i \group_i (\vecx - \fixedv_i)\\
2 y_i (\vecx - \fixedv_i)^T \group_i & y_i^2 + ||\vecx - \fixedv_i||_{\group_i}^2
\end{array}
\right]
\end{align}

Since the variable $y_i$ only appears in $\phi(\vecx, y_i)$, we may use
partial Cholesky factorization to arrive at a linear system without it.
The $n \times n$ matrix that we obtain is:

\begin{align}
& \group_i (y_i^2 - ||\vecx - \fixedv_i||_{\group_i}^2)
+ \left( 2 - 4\frac{y_i^2}{y_i^2 + ||\vecx - \fixedv_i||_{\group_i}^2} \right)
     \group_i (\vecx - \fixedv_i) (\vecx - \fixedv_i)^T \group_i \nonumber \\
= & (y_i^2 - ||\vecx - \fixedv_i||_{\group_i}^2)
\left( \group_i - \frac{2}{y_i^2 + ||\vecx - \fixedv_i||_{\group_i}^2}    \group_i (\vecx - \fixedv_i) (\vecx - \fixedv_i)^T \group_i ) \right)
\end{align}

Note that by the Cauchy-Schwarz inequality this is a PSD matrix.

Since $\phi(\vecx, \vecy) = \sum_{i} \phi(\vecx, y_i)$,
the pivoted version of $\nabla^2 \phi(\vecx, \vecy)$
can be written as:

\begin{align}
\sum_{i} \alpha_i \group_i - \beta_i \vecu_i \vecu_i^T
\end{align}

Where $\vecu_i = \group_i (\vecx - \fixedv_i)$.
To solve this linear system we invoke the Sherman-Morrison-Woodbury formula.

\begin{fact} \mbox{(Sherman--Morrison--Woodbury formula)}

If $\mata$, $\matu$, $\matc$, $\matv$ are $n \times n$, $n \times k$,
$k \times k$ and $k \times n$ matrices respectively, then:

\begin{align*}
(\mata + \matu \matc \matv)^{-1}
= \mata^{-1} -
    \mata^{-1} \matu ( \matc^{-1} + \matv \mata^{-1} \matu)^{-1} \matv \mata^{-1}
\end{align*}
\end{fact}

In our case we have $\mata = \sum_i \alpha_i \group_i$, $\matc = -\mident$,
$\matu = [\sqrt{\beta_1} \vecu_1 \ldots \sqrt{\beta_k} \vecu_k]$
and $\matv = \matu^T$.
So the linear system that we need to evaluate becomes:

\begin{align}
\mata^{-1} - \mata^{-1} \matu
( \matu^T \mata^{-1} \matu - \mident)^{-1} \matu^T \mata^{-1}
\end{align}

The system $\mata^{-1} \matu$ can be found using $k$ solves in
$\mata = \sum_i \alpha_i \group_i$, which is equivalent to the
quadratic minimization problem.
Multiplying this by $\matu$ can be done in $O(k^2n)$ time and
gives us $\matu^T \mata^{-1} \matu$.
This $k \times k$ system can in turn be solved in $k^{\omega}$ time.
The other terms can be applied to vectors in either solves in $\mata$
or matrix multiples in $\matu$, taking $O(T(n, m) + kn)$ time.
\QED

\section{Other Variants}
\label{sec:variants}

Although our formulation of $\linearprimal$ as a sum of $L_2$ objectives
differs syntactically from some common formulations, we show below that
the more common formulation involving a $L_2$-squared fidelity term can
be reduced to finding exact solutions to $\linearprimal$
using 2 iterations of ternary search.
Most other formulations differs from our formulation in the fidelity term,
but more commonly have $L_1$ smoothness terms as well.
Since the anisotropic smoothness term is a special case of the isotropic
one, our discussion of the variations will assume anisotropic objectives.

\subsection{$L_2^2$ loss term}

The most common form of the total variation objective used in practice is
one with $L_2^2$ fidelity term.
This term can be written as $||\vecx - \fixedv_0||_2^2$, which corresponds
to the norm defined by $\mident = \group_0$.
This gives:

\begin{align*}
\min_{\vecx} ~~~~~& ||\vecx-\fixedv_0||_{\group_0}^2 +
  \sum_{1 \leq i \leq k}  ||\vecx - \fixedv_i||_{\group_i}
\end{align*}

We can establish the value of $||\vecx-\fixedv_0||_{\group_0}^2$
separately by guessing it as a constraint.
Since the $t^2$ is convex in $t$, the following optimization problem is
convex in $t$ as well:
\begin{align*}
\min_{\vecx} ~~~~~& \sum_{1 \leq i \leq k}  ||\vecx - \fixedv_i||_{\group_i}
||\vecx-\fixedv_0||_{\group_0}^2
\leq t^2
\end{align*}

Also, due to the convexity of $t^2$, ternary searching on the minimizer of this
plus $t^2$ would allow us to find the optimum solution by solving
$O(\log{n})$ instances of the above problem.
Taking square root of both sides of the
$||\vecx-\fixedv_0||_{\group_0}^2 \leq t^2$
condition and taking its Lagrangian relaxation gives:

\begin{align*}
\min_{\vecx} \max_{\lambda \geq 0}
\sum_{i = 1}^k ||\vecx - \fixedv_i||_{\group_i}
    + \lambda(||\vecx - \fixedv_0||_{\group_0} - t)
\end{align*}

Which by the min-max theorem is equivalent to:

\begin{align*}
\max_{\lambda \geq 0}
-\lambda t +
\left(
    \min_{\vecx}
    \sum_{i = 1}^k  ||\vecx - \fixedv_i||_{\group_i}
        + \lambda ||\vecx - \fixedv_0||_{\group_0}
\right)
\end{align*}

The term being minimized is identical to our formulation
and its objective is convex in $\lambda$ when $\lambda \geq 0$.
Since $-\lambda t$ is linear, their sum is convex and another ternary
search on $\lambda$ suffices to optimize the overall objective.

\subsection{$L_1$ loss term}

Another common objective function to minimize is where the fidelity
term is also under $L_1$ norm. In this case the objective function
becomes:
\[
||\vecx-\fixedv_0||_1
+ \sum_{i} \sum_{1 \leq i \leq k}
    ||\vecx - \fixedv_i||_{\group_i}
\]

This can be rewritten as a special case of $\linearprimal$ as:
\[
\sum_u\sqrt{(x_u - s_u)^2}
    + \sum_{i} \sum_{1 \leq i \leq k}
        ||\vecx - \fixedv_i||_{\group_i}
\]

Which gives, instead, a grouped least squares problem with $m + k$
groups.

\end{appendix}

\end{document}